\documentclass[12pt]{article} 

\usepackage{lineno,hyperref}
\pdfoutput=1
\usepackage{graphics}
\usepackage[russian,english]{babel}
\usepackage[english] {babel}
\usepackage{amssymb,amsfonts,amsmath,mathtext}

\usepackage{graphicx}
\usepackage[font=small,skip=0pt]{caption}
\usepackage{subcaption}

\usepackage{longtable}
\usepackage{xcolor}

\usepackage{natbib}
\setcitestyle{authoryear,open={(},close={)}}

\usepackage[nottoc]{tocbibind}

\newcommand{\Ex}{{\mathsf{E}\,}}
\newcommand{\Var}{{\mathsf{Var}}\,}
\newcommand{\R}{{\mathbb R}}
\newcommand{\T}{{\mathbb T}}
\newcommand{\Z}{{\mathbb Z}}
\renewcommand{\S}{{\mathbb S}}

\newcommand{\constant}{{\mathsf{const}}}

\renewcommand{\d}{{\rm d}}

\newcommand{\eg}{e.g.\ }
\newcommand{\eeg}{E.g.\ }

\newcommand{\ie}{i.e.\ }
\newcommand{\wrt}{w.r.t.\ }
\newcommand{\rhs}{r.h.s.\ }

\newcommand{\e}{\rm e}
\renewcommand{\i}{{\mathsf i}}
\modulolinenumbers[5]

   \title{A spatio-temporal stochastic pattern generator\\
   for simulation of uncertainties in geophysical ensemble prediction and 
   ensemble data assimilation\footnote{
   This manuscript is an extended version of the article \citep{TsyrulnikovGayfulin_MZ}.
   In particular, Appendices A,B, and D in this manuscript are missing in the article.
   But a description of an implementation of the Stochastic
   Pattern Generator in the meteorological 
   COSMO model is present in the article but is missing here.   
   }
   }
   \author{Michael Tsyrulnikov  and Dmitry Gayfulin\\
   \smallskip
   {\em HydroMetCenter of Russia}\\
   {\small (mik.tsyrulnikov@gmail.com) }}
   
\begin{document}

\maketitle


\begin{abstract}

A generator of spatio-temporal pseudo-random Gaussian fields that satisfy the 
``proportionality of scales'' property (Tsyroulnikov, 2001)  is presented.
The generator is based on a third-order in time stochastic  
differential equation with a pseudo-differential spatial operator defined on a 
limited area 2D or 3D domain in the Cartesian coordinate system.
The generated pseudo-random fields are homogeneous (stationary) and isotropic in space-time
(with the scaled vertical and temporal coordinates).
The correlation functions in any spatio-temporal direction belong to the Mat\'ern class.
The spatio-temporal correlations are non-separable.
A spectral-space numerical solver is implemented and accelerated exploiting
properties of real-world geophysical fields, in particular, smoothness of their spatial spectra.
The generator is designed to create additive or multiplicative, or other spatio-temporal perturbations
that represent uncertainties in  numerical prediction models in geophysics.
The program code of the generator is publicly available.

\end{abstract}


\tableofcontents

\section {Introduction}

\subsection{Stochastic dynamic prediction}
\label{sec_intro_stochdyn}

Since the works of 
\citet{Epstein} and
\citet{Tatarsky}, 
we know that accounting for the uncertainty in the initial forecast fields 
can improve weather (and other geophysical) predictions.
Assigning a probability distribution for the truth at the start of the forecast
(instead of using  deterministic initial data) and attempting to advance this distribution in time
according to the dynamic (forecast) model is called stochastic dynamic prediction.

The advantage of the stochastic dynamic prediction paradigm is twofold.
First, the resulting forecast probability distribution provides a valuable measure of the {\em uncertainty}
in the prediction, leading to probabilistic forecasting 
and flow-dependent background-error statistics in data assimilation.
Second, for a nonlinear physical model, switching from the deterministic forecast
to the mean of the forecast probability distribution
improves the mean-square accuracy of the prediction, \ie can improve the deterministic forecasting.

\subsection{Model errors}
\label{sec_intro_mem}

Since \citet[][]{Pitcher}, we realize that not only uncertainties in the initial data
(analysis errors) matter, forecast model (including boundary conditions) imperfections 
also play an important role.
Simulation of model errors is the subject of this study, so we define them now.
Let the forecast model be of the form
\begin{linenomath*}
\begin {equation}
\label {m}
\frac {\d {\bf x}}{\d t} = {\bf F}({\bf x}),
\end {equation}
\end{linenomath*}
where $t$ is time, ${\bf x}$ is the vector that represents the (discretized) state of the system,
and ${\bf F}$ is the model (forecast) operator.
The imperfection of the model Eq.(\ref{m}) means that the (appropriately discretized) 
truth does {\em not} exactly satisfy this equation. The discrepancy is called the model error \citep[e.g.][]{Orrell}:
\begin{linenomath*}
\begin {equation}
\label {me}
\boldsymbol\xi^t =  {\bf F}({\bf x}^t) - \frac {\d {\bf x}^t}{\d t}.
\end {equation}
\end{linenomath*}
The true model error $\boldsymbol\xi^t$ is normally unknown.
In order to include model errors in the stochastic dynamic prediction paradigm, 
one {\em models} $\boldsymbol\xi^t(t)$ as a {\em random process}, $\boldsymbol\xi(t)$, or, 
in other words, as a spatio-temporal random field $\xi(t, {\bf s})$ 
(where ${\bf s}$ is the spatial vector).
The probability distribution of $\boldsymbol\xi(t)$ 
(in most cases, dependent on the flow)
is assumed to be known.

Rearranging the terms in Eq.(\ref{me}),
and replacing the unknown $\boldsymbol\xi^t$ with its stochastic counterpart 
$\boldsymbol\xi$,
  we realize that the resulting model of
truth is the stochastic dynamic equation 
\begin{linenomath*}
\begin {equation}
\label {msd}
\frac {\d {\bf x}}{\d t}=  {\bf F}({\bf x}) - \boldsymbol\xi.
\end {equation}
\end{linenomath*}

Thus, the extended stochastic dynamic prediction (or modeling) paradigm requires two input probability
distributions (that of initial errors and that of model errors) and aims to 
transform them to the output (forecast) probability distribution.

\subsection{Ensemble prediction}
\label{sec_intro_ensm}

Stochastic dynamic modeling of complex geophysical systems
is hampered by their high dimensionality and non-linearity.
For realistic models, the output probability distribution is analytically 
intractable. An affordable approximate solution is provided 
by the Monte-Carlo method called in geosciences {\em ensemble prediction}.

In ensemble prediction, the input  uncertainties (\ie initial  and model errors)
are represented by simulated {\em pseudo-random draws}
from the respective probability distributions. 
A relatively small affordable number of these draws are fed to the forecast model 
giving rise to an {\em ensemble} of predictions (forecasts). 
{\em If initial  and model errors are sampled from the correct respective distributions, 
then the resulting forecast ensemble members are draws from the correct probability distribution
of the truth given all available external data (initial and boundary conditions).}
This mathematically justifies the ensemble prediction principle.
From the practical perspective, members of the forecast ensemble can be interpreted as ``potential
truths'' consistent with all available information.

In what follows, we concentrate on the model error field $\xi(t, {\bf s})$.
We briefly review existing models for $\xi(t, {\bf s})$ and then present our 
stochastic pattern generator, whose goal is to simulate pseudo-random draws of $\xi(t, {\bf s})$
from a meaningful and flexible distribution.

\subsection{Practical model error modeling}
\label{sec_intro_pmem}

In meteorology, our knowledge of the actual model error probability distribution  is scarce.
Justified stochastic model-error models are still to be devised and verified.
In the authors' opinion, the best way to stochastically represent spatio-temporal
forecast-model-error fields is to treat each error source separately,
so that, say, each physical parametrization is accompanied with a spatio-temporal 
stochastic model of its uncertainty. Or, even better, to completely switch 
from deterministic physical parameterizations to stochastic ones.
There is a growing number of such developments \citep[see][for a review]{Berner}, but the problem is 
so complex that we cannot expect it to be solved in the near future.
Its solution is further hampered by the fact that the existing 
meteorological observations are too scarce and too inaccurate
for model errors to be objectively identified by comparison with measurement data
with satisfactory accuracy \citep{TsyrulnikovGorin}.

As a result, in meteorology, {\em ad-hoc} model-error models are in wide use. The  existing
approaches can be classified as either non-stochastic or stochastic.
Non-stochastic schemes can be multi-model (different ensemble members are generated using 
different forecast models) or multi-parameterization (each ensemble member is generated using
the forecast model with a unique  combination of different physical parameterization schemes or their
parameters). 
These techniques are capable of introducing significant diversity in the ensemble \citep{Berner2011},
but the resulting ensemble members cannot be considered as independent and drawn from the 
same probability distribution (an assumption normally made in using the ensembles).
Besides, there are not enough different models and not enough substantially different
physical parameterizations to generate large  ensembles.
Finally, running many forecast models is a technologically very demanding task.

Stochastic approaches, on the contrary, offer the opportunity to generate as many
ensemble members taken from the same probability distribution as needed, while working with just one
forecast model and one set of physical parameterizations.
In atmospheric ensemble prediction and ensemble data assimilation, 
the most widely used stochastic techniques are 
SPPT (Stochastic Perturbations of Physical Tendencies, \citet[][]{Buizza}),
SKEB (Stochastic Kinetic Energy Backscatter scheme, \citet{Shutts}), and
SPP (Stochastically Perturbed Parameterizations, \citet{Christensen,Ollinaho}).
In the SPPT, multiplicative perturbations to the {\em tendencies} produced by the model's
physical parameterizations are introduced. The multiplier is a spatio-temporal random field
centered at 1.
In the SKEB, {\em additive} perturbations are computed by modulating a spatio-temporal random
field by the local kinetic energy dissipation rate.
In the SPP, selected {\em parameters} of the  physical parameterization schemes are perturbed
again using a spatio-temporal field, which thus is seen to be needed in 
all of the above stochastic model error representation schemes.
Stochastic parameterization schemes can also demand such fields \citep[e.g.][]{Bengtsson}.

\subsection{Generation of spatio-temporal random  fields}
\label{sec_intro_gen}

The simplest non-constant pseudo-random field is the white noise, \ie
the uncorrelated in space and time random field. The white noise is the 
default forcing in stochastic differential equations, \eg \citet{Jazwinski} or \citet{Arnold}.
Its advantage is the complete absence of any spatio-temporal structure, it is a pristine 
source of stochasticity. But in model-error modeling, this lack of structure 
precludes its direct use as an additive or multiplicative perturbation field
because model errors are related to the weather pattern and so should be correlated
(dependent) both in space and time.
\citet{Tsyrulnikov2005} showed in a simulation study that model errors can exhibit complicated 
spatio-temporal behavior.

A correlated pseudo-random spatio-temporal field can be easily computed
by generating independent random numbers at  points of a {\em coarse} spatio-temporal grid
and then assigning each of them to all model  grid  points within the 
respective coarse-grid cell \citep{Buizza}. 
As a  result, the model-grid field
becomes correlated in space and time. The decorrelation space and time scales are, obviously,
defined by the respective coarse grid spacings
(\eg in \citet{Buizza} these were  about 1000 km in space and 6 h in time).
This technique is extremely simple but it suffers from two flaws.

First, the resulting model-grid field appears to be discontinuous and inhomogeneous.
Second, the spatio-temporal structure of the 
field is not scale dependent, that is, the resulting temporal
length scales do not depend on the respective spatial scales.
In reality, longer spatial scales ``live longer'' than shorter spatial scales,
which ``die out'' quicker. This `proportionality of scales' is widespread 
in geophysical fields \citep[see][and references therein]{Tsyroulnikov2001}
and other media, \citep[e.g.][p.129]{Meunier},
so we believe this property should be represented by model-error models. 
Note also that the ``proportionality of scales'' is a special case of 
the {\em non-separability} of
spatio-temporal covariances. For a critique of simplistic separable space-time covariance models,
see \citet{CressieHuang}, \citet{Stein2005}, \citet{Gneiting}, and section \ref{sec_req} below.

Another popular space-time pseudo-random field generation technique 
employs a spectral transform in space and then imposes independent
temporal auto-regressions for the coefficients of the spectral expansion 
\citep{Berner2009,Palmer,Charron,Bouttier2012}.
This technique is more general and produces homogeneous fields, but the above 
implementations use {\em the same time scale} for all spatial wavenumbers
so that there are still no space-time interactions in the generated spatio-temporal fields
(though \citet{Charron} noted that the decorrelation time scales can be made dependent on the 
spatial scales and \citet{Palmer} allowed for this dependence in their SKEB pattern generator equations).

In this  report, we propose and test a spatio-temporal 
Stochastic (pseudo-random) Pattern Generator (SPG)
that accounts for the above ``proportionality of scales'' and 
imposes meaningful space-time interactions. 
The SPG operates on a limited-area domain.
It is based on a (spectral-space) solution to a stochastic partial differential equation,
more precisely, to a stochastic  differential equation in time with a pseudo-differential spatial operator.
In what follows, we present the technique, 
examine properties of the resulting spatio-temporal fields on 2D and 3D spatial domains,
describe the numerical scheme, and 
explore the performance of the SPG.
The technique is implemented as a Fortran program freely available from 
\url{https://github.com/gayfulin/SPG}.

\section {Model error  fields: separability vs. ``proportionality of scales''}
\label{sec_req}

In this motivational section we show on a simple 1D (in space) example that 
space-time interactions in the model error random field  play a significant role.
Specifically, we demonstrate that these interactions determine whether the 
spatial length scale of the resulting forecast error field grows, in a first
approximation,  in time or remains constant.

We note that for small enough model error perturbations and small enough lead times,
the forecast error due to the accumulated model errors can be approximated by the so-called 
model-error drift, that is, the time integrated model error:
$\bar\xi(t,s) =\int_0^t \xi(t,s) \,\d t$  \citep[][]{Orrell}.
Therefore, the methodology in this  section is to take two fields,
one with separable spatio-temporal correlations and the other 
with``proportional scales'', integrate them in time, and look at
the spatial length scales of the two time integrated random fields.

Theoretically, the time integration reduces (filters out) small-scale-in-time 
components of the  field. As a separable field has no space-time interactions, its time integral
should have exactly the same spatial length scale as $\xi(t,s)$.
For a proportional-scales field,  smaller scales in time are associated with  
smaller scales in space, so the amount of small spatial scales in the time integrated 
field should decrease in time leading to an increase in the spatial length scale.

To verify these theoretical conclusions, we set up the following numerical experiment.
We considered a 1D domain of size 100 km and the time integration period of 3 h.
In this 2D spatio-temporal domain, we introduced a grid  
with 100 points in space and 100 points in time.
On this grid, we simulated two random fields, both with unit variance and exactly the same
spatial and temporal exponential correlations. 
The first field had separable correlations 
$C_1(\Delta t,\Delta s)=\exp(-|\Delta s|/L) \cdot \exp(-|\Delta t|/T)$,
whereas the second field had non-separable correlations
$C_2(\Delta t,\Delta s)=\exp(-\sqrt{(\Delta s/L)^2 + (\Delta t/T)^2}$,
which can be shown to satisfy the ``proportionality of scales'' property.
%
%
The spatial length scale $L$ was selected in such a way that the spatial correlation
function intersects the 0.7 level at the distance of 50 km. 
The temporal length scale was selected to be equal to $L/U$, where $U=20$ m/s was
taken as the characteristic flow velocity.
Note that both the separability and the exponential temporal correlation function are what the
scale-independent  first-order auto-regressions
used in \citet{Berner2009}, \citet{Palmer}, \citet{Charron}, and \citet{Bouttier2012} imply.

\begin{figure}
\begin{center}
   { \scalebox{0.25}{ \includegraphics{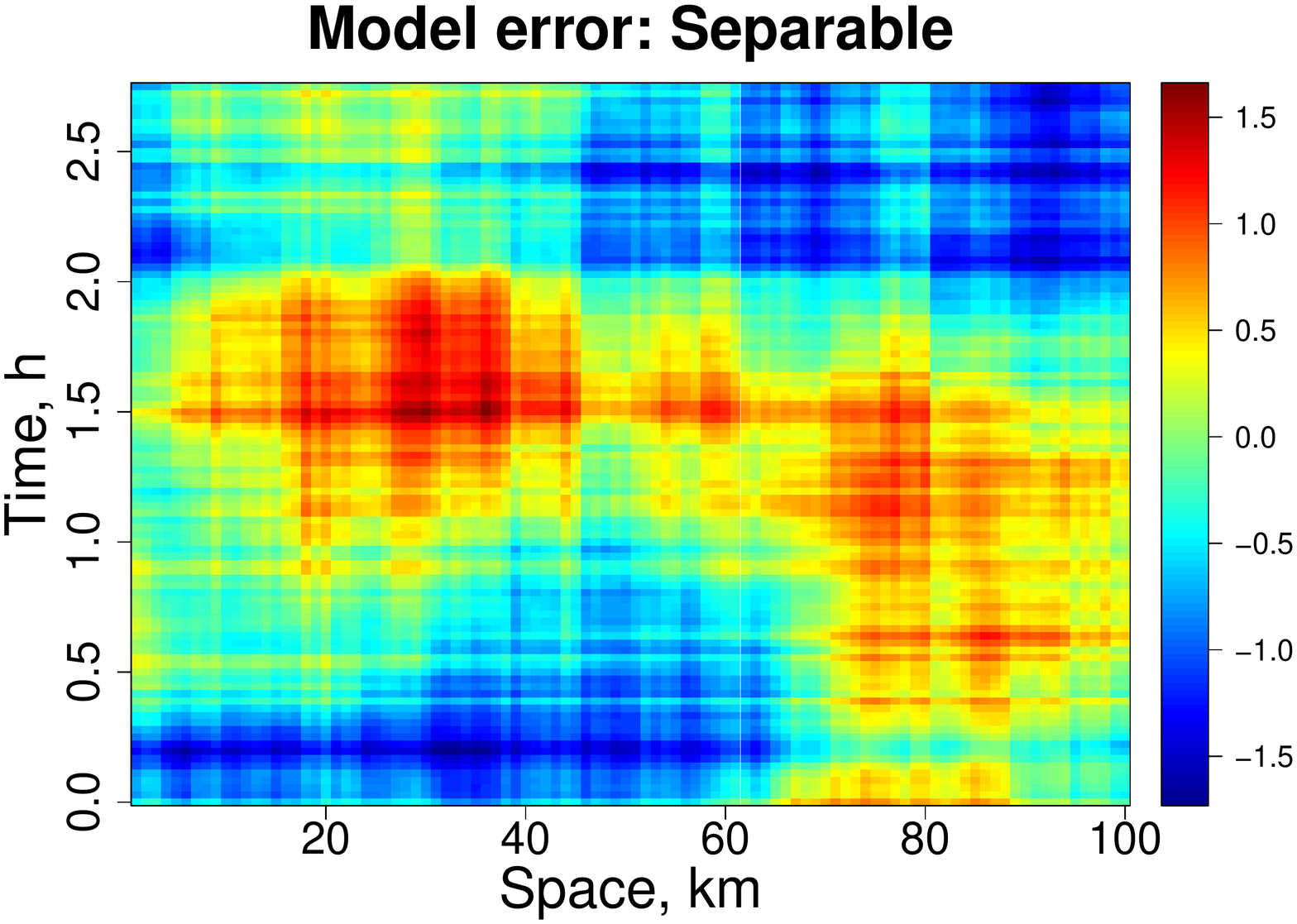}}
     \scalebox{0.25}{ \includegraphics{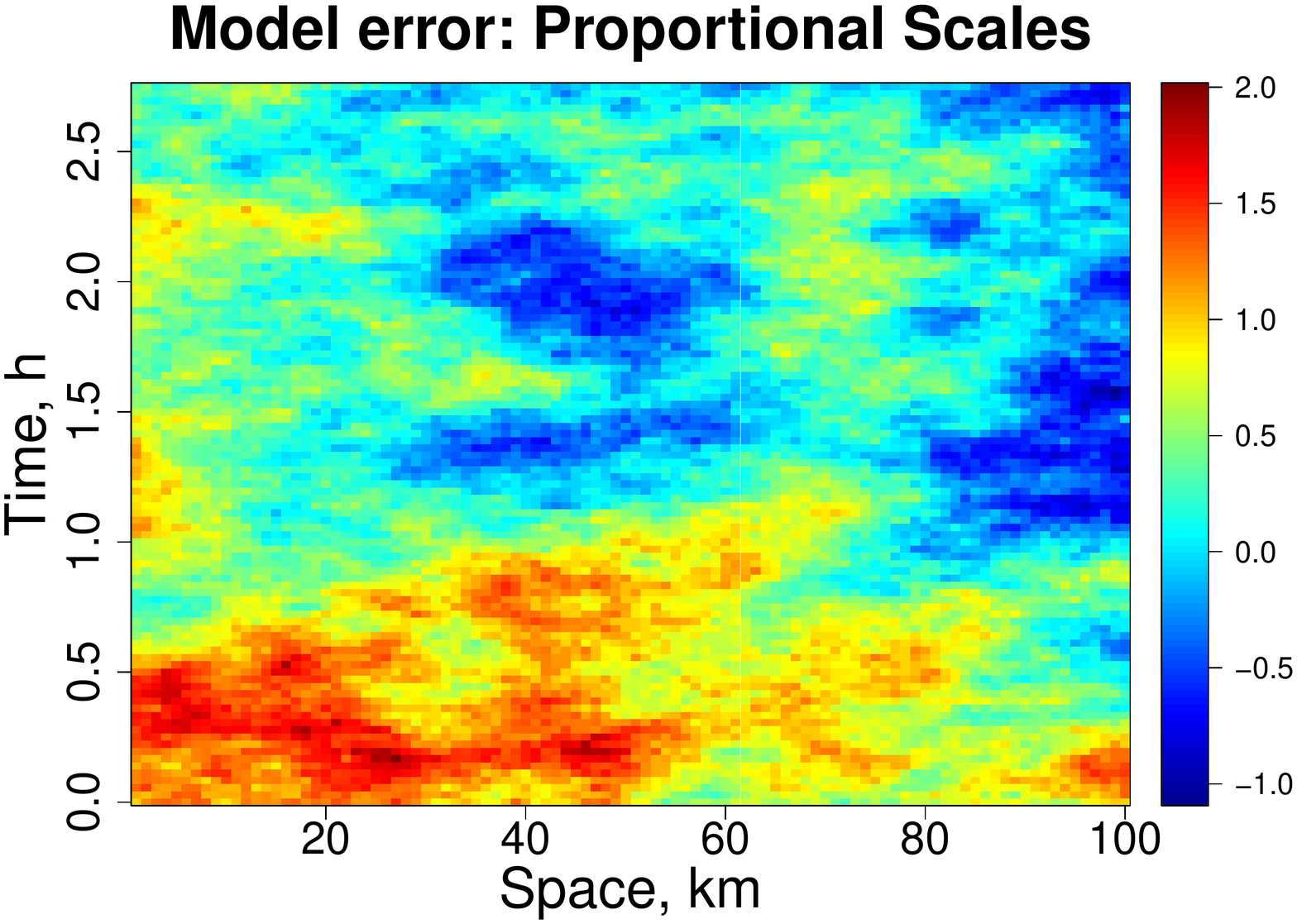}}
       }
\end{center}
  \caption{Simulated spatio-temporal fields. 
  {\em Left}::  With separable space-time correlations. 
  {\em Right}:  With  non-separable proportional-scales correlations.
  }
\label{Fig_ME}
\end{figure}

Knowing the two correlation functions, we simulated pseudo-random realizations of the 
two fields (by building the two covariance matrices, computing their square roots, and
applying the latter to vectors of independent $N(0,1)$ random variables), see Fig.\ref{Fig_ME}.
Comparing the two panels of Fig.\ref{Fig_ME}, one can see that the two fields look  quite differently.
Visually, the most striking difference is the lack of isotropy in the separable case.
The proportional-scales field looks much more realistic than the separable one.

\begin{figure}
\begin{center}
   { \scalebox{0.34}{\includegraphics{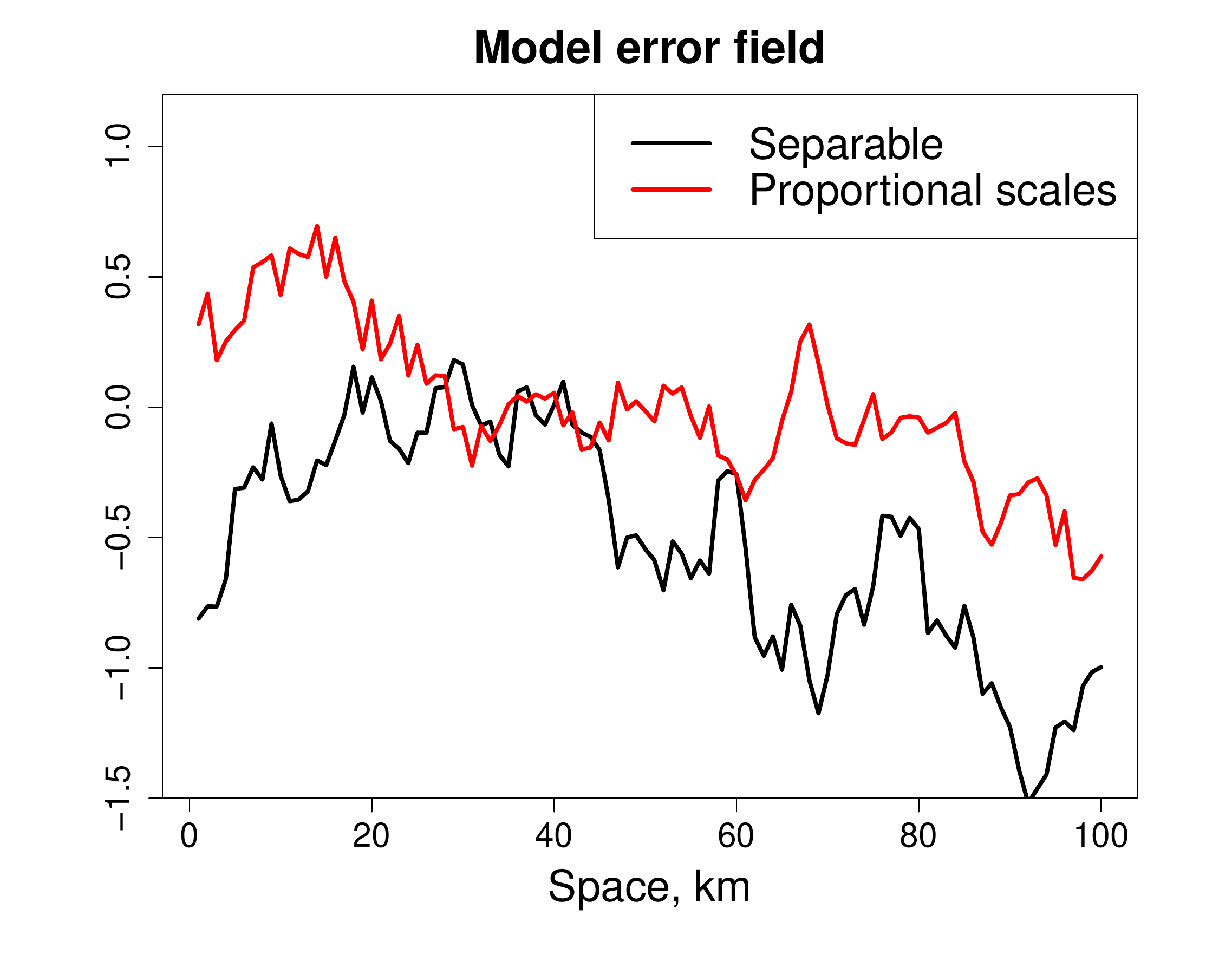}}
     \scalebox{0.34}{\includegraphics{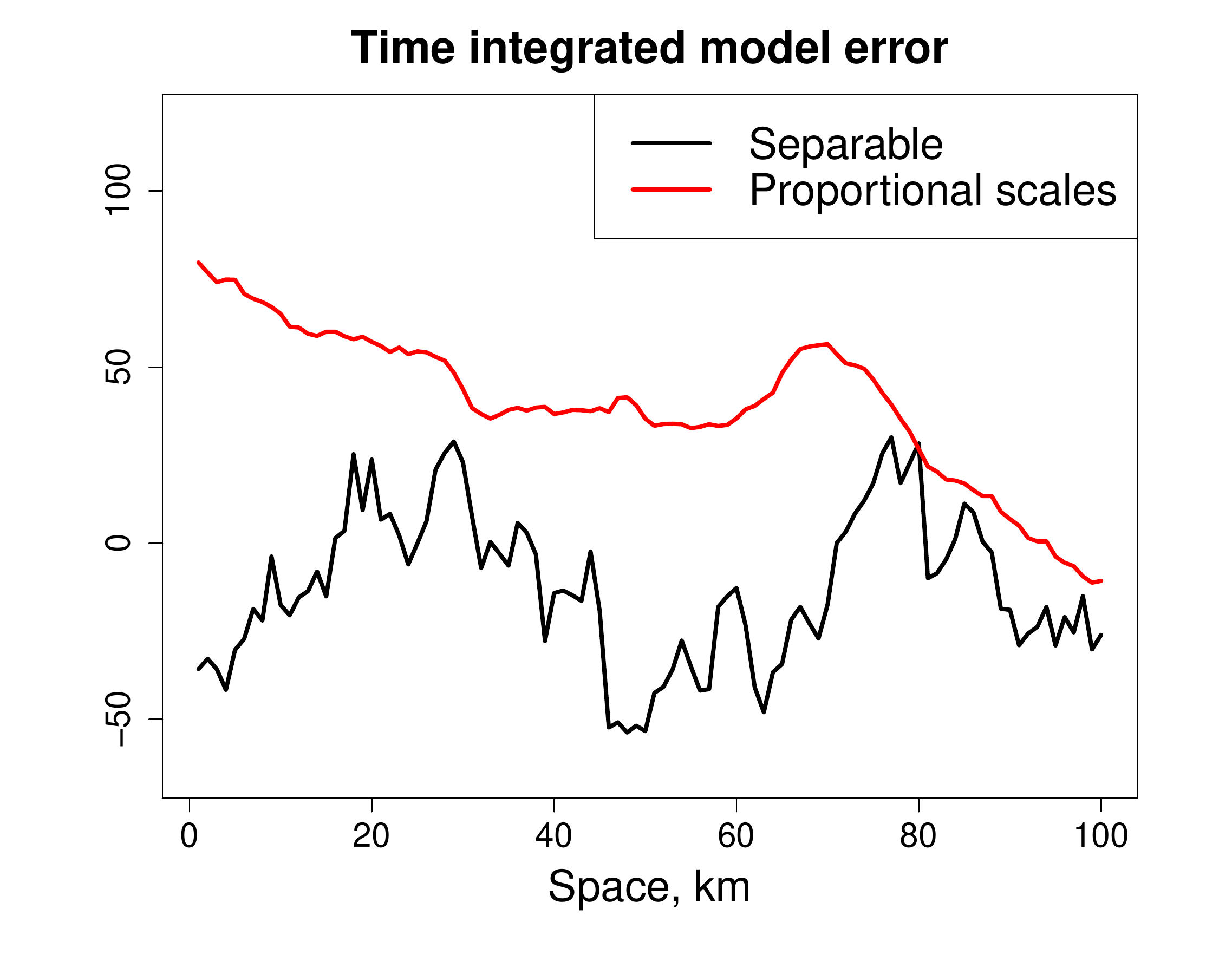}}
       }
\end{center}
  \caption{Spatial cross-sections of the simulated fields. 
  {\em Left}::  Model error fields. 
  {\em Right}:  Time integrated model error fields.
  }
\label{Fig_MEFE}
\end{figure}

To get a more objective criterion, we computed the time integrated model error field $\bar\xi(t,s)$
(the model error drift, a proxy to the model-error induced forecast error, 
see above in this section).
Figure \ref{Fig_MEFE} shows the spatial cross-sections of the 
arbitrarily chosen realizations of the  model error fields  (left) and 
the drift fields (right).
The realizations generated by the separable random field model are given in black
and the realizations of the proportional-scales field are represented by the red curves.
One can see that, indeed, the time integration did not change the spatial structure of the 
{\em separable} field (compare the two black curves in   Fig.\ref{Fig_MEFE}, left and right).
In contrast, the time integrated {\em proportional-scales} field becomes much smoother in space 
(compare the two red curves in   Fig.\ref{Fig_MEFE}, left and right).

\begin{figure}
\begin{center}
   { \scalebox{0.35}{ \includegraphics{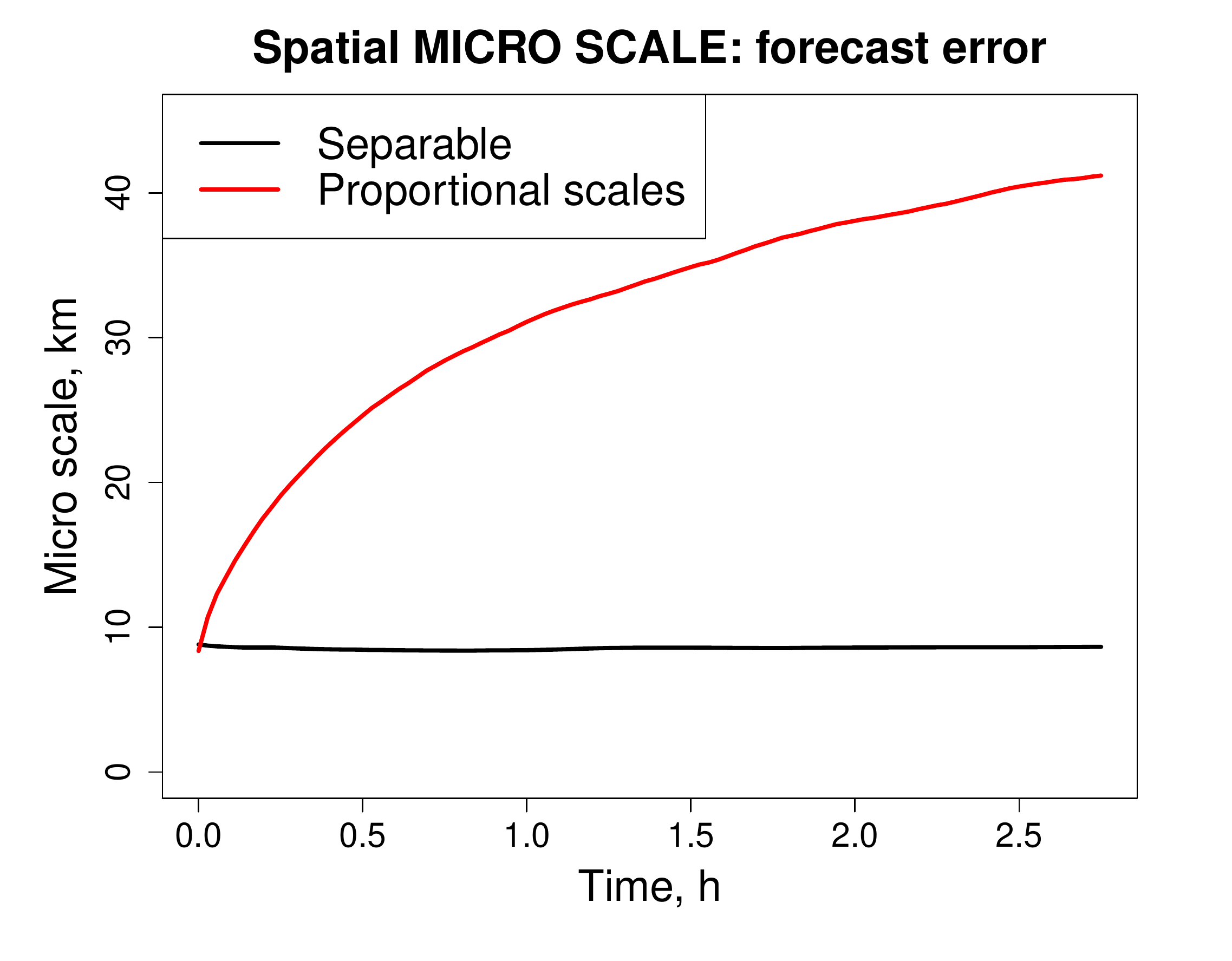}}
            }
\end{center}
  \caption{Spatial micro-scale as a function of integration time
  for the two time integrated random fields (separable and proportional-scales).
  }
\label{Fig_Lx}
\end{figure}

Even more objectively, we estimated the {\em spatial micro-scale} 
of the drift $\bar\xi(t,s)$. 
The estimator was 
$({ \Var \bar\xi} / {\Var \delta\bar\xi})^{1/2}\cdot h$,
where 
$\delta\xi$ is the forward finite difference in space,  
the variance $\Var$ was estimated by averaging over the space coordinate 
and over an ensemble of 100 realizations), and
$h$ is the spatial mesh size.
The resulting spatial micro-scales for the two
fields in question are displayed in Fig.\ref{Fig_Lx}
as functions of time.
As expected, in the separable case, the spatial micro-scale  did not change 
as a result of the time integration (the flat black line), whereas in the non-separable 
proportional-scale case, the spatial length scale of the model error drift
rapidly grew  in time.
It is worth emphasizing that in data assimilation,
the spatial length scale is a very important
attribute of the forecast error field and thus needs to be correctly represented by a
forecast (background) ensemble.

Thus, we have shown that the specific type of the spatio-temporal interactions in a model (tendency) error
field has important consequences for the spatial structure of the 
resulting practically relevant forecast error field.
We have no evidence on the actual model error spatio-temporal structure, but we know
that non-separability and, more specifically, proportionality of scales is ubiquitous
in geophysics  \citep{Tsyroulnikov2001}, therefore,
 we postulate that the SPG should produce proportional-scales fields.

\section {SPG: Requirements and approach}
\label{sec_req}

The general requirements are:

\begin{enumerate}
\item
\label{req_SPG_isotr}
The SPG should produce univariate stationary in time and
homogeneous (stationary) and isotropic in space Gaussian pseudo-random fields $\xi(t, {\bf s})$
in 3D and 2D spatial domains. 
\item
The SPG should be fast enough so that it does not significantly slow down 
the  forecast model computations. 
\item
The magnitude as well as the spatial and temporal length scales of $\xi(t, {\bf s})$ are to be tunable. 
\end{enumerate}

\noindent We also impose more specific  requirements:
\begin{enumerate}
\setcounter{enumi}{3}
  \item
  \label{req_SPG_cont}
The random field  $\xi(t, {\bf s})$ should have 
finite variance and  continuous realizations (sample paths).
 \item
 \label{req_SPG_PS}
The spatio-temporal covariances should obey the ``proportionality of scales'' principle: 
larger (shorter) spatial scales should be associated with larger (shorter)  temporal scales
\citep{Tsyroulnikov2001}.
\item
\label{req_SPG_flex}
The SPG {\em ansatz} should be
flexible enough to allow
for practicable solutions in both physical space and spectral space.
\end{enumerate}

Two comments are in order. 
Firstly,  stationarity, homogeneity,  isotropy, and Gaussianity imposed by requirement \ref{req_SPG_isotr}
are just the simplest natural properties of a spatio-temporal random field.
The SPG is intended to  be used as a building block
in practical schemes like the above SPPT, SKEB, SPP, or others.
Its role is to be the source of meaningful and easily tunable spatio-temporal stochasticity, 
whereas physical model error features (flow dependence, non-Gaussianity, etc.)
are to be provided by the specific model error modeling scheme on a point-by-point basis.   

Secondly, requirement \ref{req_SPG_flex} demands an SPG equation to be solvable in physical space as well as in spectral space
for the following reasons. All the above mentioned existing pattern generators are 
spectral space based because this is the simplest way to get a homogeneous and isotropic 
field in physical space. So, following this path, we would like to have a spectral-space solver.
But we envision that a combination of a homogeneous and isotropic spatial structure 
(provided by the SPG) and point-by-point flow dependent and/or non-Gaussian features
(provided by the specific model error modeling scheme)
can appear too restrictive in the near future. Specifically, this
combination approach cannot produce  variable local spatial and temporal length scales
if used in schemes like the SKEB or SPP (say, we may wish to reduce the local 
length scales in meteorologically active areas like cyclones or convective systems).
Therefore, we wish the SPG equation to allow for a physical-space
solver that would be capable of imposing variable in space and time structures.

As a starting point in the development of the SPG, 
we select the general class of {\em linear  evolutionary  stochastic
partial differential equations} (SPDE).
This choice is motivated by the  flexibility of this class of spatio-temporal models
\citep[e.g.][]{Lindgren2011}.
In particular, for an SPDE, it is relatively easy to introduce inhomogeneity  
(non-stationarity) in space and time as well as
local anisotropy---either by changing coefficients of the spatial operator
or by changing local properties of the driving noise. 
One can also produce non-Gaussian fields by making the random forcing
non-Gaussian \citep[e.g.][]{Aberg,Wallin}. Physical-space discretizations of 
SPDEs lead to {\em sparse} matrices, which give rise to  fast numerical algorithms.
If an SPDE has constant coefficients, then it can be efficiently solved 
using spatial  spectral-space expansions. 

In this study, we develop the SPG  that relies 
on a spatio-temporal stochastic model with {\em constant coefficients} so that both physical-space
and spectral-space solvers can be employed. 
To facilitate the spectral-space solution, the general strategy is to define the SPG model on a standardized spatial domain. 
The operational pseudo-random fields are then produced by mapping of the generated fields
from the standardized  domain to the forecast-model domain.
In 3D, the standardized spatial domain is chosen to be  the  unit cube with the periodic
boundary conditions in all three dimensions, in other words,
 the  three-dimensional (3D) unit  torus.
In 2D,  the standardized domain is the 2D unit torus.
The 3D and 2D cases are distinguished 
by the dimensionality $d=2$ or $d=3$ in what follows.
To simplify the presentation, the default dimensionality will be $d=3$.

\section {Tentative first-order SPG model}
\label{sec_order1}

\subsection{Physical-space model}
\label{sec_order1_phy}

The random field in question $\xi(t,{\bf s})$ is a function of the time coordinate $t$ 
and the space vector ${\bf s}=(x,y,z)$, where $(x,y,z)$ are the three spatial coordinates.
Each of the spatial coordinates belongs to the the unit circle $\S^1$,
so that  
${\bf s}$ is on the unit torus
$T^3=\S^1 \times \S^1 \times \S^1$ ($\T^2$ in the 2D case).

We start with the simplest general form of the first-order Markov model:
\begin{linenomath*}
\begin {equation}
\label {spg_gen}
\frac {\partial \xi(t,{\bf s})}{\partial t} + A \, \xi(t,{\bf s}) =  \alpha(t,{\bf s}),
\end {equation}
\end{linenomath*}
where $A$ is the spatial linear operator to be specified and $\alpha$ is the driving noise.
$\alpha$ is postulated to be white in space and time; this is done to facilitate a fast numerical solver
in physical space as demanded by requirement \ref{req_SPG_flex} 
in section \ref{sec_req} because generation of the white noise 
is computationally inexpensive (its values on a grid in space and time are just independent
Gaussian random variables).

The SPG is required to be fast, so we choose  $A$ to be a {\em differential} operator
(because, as we noted, in this case a physical-space discretization of $A$  gives rise to
a very sparse matrix).

Further, since we wish $\xi(t,{\bf s})$ to be homogeneous and isotropic in space,
we define $A$ to be a polynomial of the negated spatial Laplacian:
\begin{linenomath*}
\begin {equation}
\label {ALapl}
A = P(-\Delta) = \sum^q_{j=0} c_j (-\Delta)^j,
\end {equation}
\end{linenomath*}
where $P(x)$ is the polynomial and $q$ its degree (a positive integer).
We will refer to $q$ as the spatial order of the SPG model.
Note that the negation of the Laplacian is convenient because  $(-\Delta)$ is 
a non-negative definite operator.

The model Eq.(\ref{ALapl}) appears to be too rich for the purposes
of the SPG at the moment, so
in what follows we employ an even more reduced (but still quite flexible) form 
\begin{linenomath*}
\begin {equation}
\label {ALapl2}
 A=P(-\Delta) = \mu (1 - \lambda^2\Delta)^q,
\end {equation}
\end{linenomath*}
where $\mu$ and $\lambda$ are positive real parameters.
So, we start with the following SPG equation:
\begin{linenomath*}
\begin {equation}
\label {bpg1}
\frac {\partial \xi(t,{\bf s})}{\partial t} +  \mu (1 - \lambda^2\Delta)^q \,\xi(t,{\bf s}) =  \alpha(t,{\bf s}).
\end {equation}
\end{linenomath*}
%

\subsection{Spectral-space model}
\label{sec_order1_spe}

On the torus $\T^\mathnormal{d}$, a Fourier series is an expansion in the basis functions 
$\e^{\i ({\bf k}, {\bf s})} \equiv \e^{\i (\mathnormal{mx+ny+lz})}$, 
where the wavevector ${\bf k}$ is, for $d=3$, the triple
of integer wavenumbers, ${\bf k}=(m,n,l)$.
We perform the Fourier decomposition for both $\alpha(t, {\bf s})$ and $\xi(t, {\bf s})$,
\begin{linenomath*}
\begin {equation}
\label {bpg7a}
\alpha(t, {\bf s})=\sum_{{\bf k} \in \Z^d} \tilde\alpha_{\bf k}(t) \e^{\i ({\bf k}, {\bf s})}
\end {equation}
\end{linenomath*}
and
\begin{linenomath*}
\begin {equation}
\label {bpg7}
\xi(t, {\bf s})=\sum_{{\bf k} \in \Z^d} \tilde\xi_{\bf k}(t) \e^{\i ({\bf k}, {\bf s})}
\end {equation}
\end{linenomath*}
(where $\Z$ denotes the set of integer numbers) 
and substitute these expansions into Eq.(\ref{bpg1}).
From the orthogonality of the basis functions, 
 we obtain that Eq.(\ref{bpg1}) decouples into the set of ordinary stochastic differential
 equations  \citep[OSDE, e.g.][]{Jazwinski,Arnold} in time:
\begin{linenomath*}
\begin {equation}
\label {osde1}
\frac{\d \tilde\xi_{\bf k}} {\d t} +  \mu (1 + \lambda^2 {k}^2)^q 
         \, \tilde\xi_{\bf k}(t)= \tilde\alpha_{\bf k}(t),
\end {equation}
\end{linenomath*}
where $k=|{\bf k}|=\sqrt{m^2+n^2+l^2}$. 
The white driving noise  $\alpha$  is stationary, hence the
spectral-space coefficients  $\tilde\alpha_{\bf k}(t)$ are 
probabilistically independent random processes. 
This is well known for random fields on the $d$-dimensional real space $\R^d$ (where spectra are continuous),
see \eg  Chapter 2 in \citet{Adler} or  section 8 in \citet{Yaglom},
and can be directly verified in our case of the fields on the torus (where spectra are discrete). 
Therefore, for different wavevectors ${\bf k}$,
the resulting spectral-space equations
 are probabilistically completely {\em independent} from each other.
This greatly simplifies the solution of the SPG equations because instead of handling
 the complicated SPDE Eq.(\ref{bpg1}) we have to solve a number of independent simple OSDEs
 Eq.(\ref{osde1}).

Further, from the postulated whiteness of the spatio-temporal random field 
$\alpha(t,{\bf s})$, all $\tilde\alpha_{\bf k}(t)$
are white in time random processes with the same intensity $\sigma$,
(Appendix \ref{app_wnoise}):
\begin{linenomath*}
\begin {equation}
\label {swn1}
\tilde\alpha_{\bf k}(t) = \sigma \, \Omega_{\bf k}(t),
\end {equation}
\end{linenomath*}
where $\Omega_{\bf k}(t)$ are the independent {\em standard} white noises,
\ie the derivatives of the independent standard Wiener processes $W_{\bf k}(t)$
such that 
\begin{linenomath*}
\begin {equation}
\label {wp1}
\Omega_{\bf k}(t) \d t = \d W_{\bf k}(t).
\end {equation}
\end{linenomath*}
Thus, the first-order SPG model reduces to a series of  OSDEs
\begin{linenomath*}
\begin {equation}
\label {osde1a}
\d \tilde\xi_{\bf k}  +  \mu (1 + \lambda^2 {k}^2)^q  \, \tilde\xi_{\bf k} \, {\d t} =
         \sigma \, \d W_{\bf k}.
\end {equation}
\end{linenomath*}
For practical purposes the series  is truncated, so that  ${\bf k}\equiv (m,n,l)$ is limited:
 $|m| < m_{\rm max}$, 
  $|n| < n_{\rm max}$, and
   $|l| < l_{\rm max}$,
where  $m_{\rm max}$, $n_{\rm max}$, and $l_{\rm max}$ are the truncation limits.
If not otherwise stated, all the truncation limits are the same and denoted by $n_{\rm max}$.

\subsection{Stationary spectral-space statistics}
\label{sec_statio}

Equation (\ref{osde1a}) is a first-order OSDE with constant coefficients sometimes
called the Langevin equation
(\eg \citet{Arnold} or \citet{Jazwinski}, Example 4.12). Its generic form is
\begin{linenomath*}
\begin {equation}
\label {osde2}
\d \eta + a \eta \,\d t= \sigma \d W,
\end {equation}
\end{linenomath*}
where  $\eta(t)$ is the random process in question, $a$ and $\sigma$ are constants, and 
$W(t)$ is the standard Wiener process.
The solution to Eq.(\ref{osde2}) is known as the Ornstein-Uhlenbeck random process,
whose stationary  (steady-state) temporal covariance function  is 
\begin{linenomath*}
\begin {equation}
\label {osde4}
B_\eta(t)=\frac{\sigma^2}{2a} \, \e^\mathnormal{-a|t|}
\end {equation}
\end{linenomath*}
\citep[e.g.][Example 4.12]{Jazwinski}. From Eq.(\ref{osde4}), it is clear that $a$ has the 
meaning of the inverse  temporal length scale $\tau = 1/a$.

Now, consider 
the stationary covariance function of the elementary random process  $\tilde\xi_{\bf k}(t)$,
\begin{linenomath*}
\begin {equation}
\label {bt}
\Ex \tilde\xi_{\bf k}(t_0) \cdot \tilde\xi_{\bf k}(t_0 +t) = b_{\bf k} \cdot C_{\bf k}(t),
\end {equation}
\end{linenomath*}
where $b_{\bf k}$ is the variance and $C_{\bf k}(t)$  the  correlation function.
According to Eq.(\ref{bpg7}), $\tilde\xi_{\bf k}$ is the spatial spectral component of the 
random field in question $\xi(t,{\bf s})$. Therefore $b_{\bf k}=\Var\tilde\xi_{\bf k}$
is called the {\em spatial spectrum} of $\xi(t,{\bf s})$.
From  Eqs.(\ref{osde1a}) and (\ref{osde4}), we have
\begin{linenomath*}
\begin {equation}
\label {bk}
 b_{\bf k} =  \frac {\sigma^2}  {2 \mu (1 + \lambda^2 {k}^2)^q}
\end {equation}
\end{linenomath*}
and $C_{\bf k}(t) = \exp( -{|t|} / {\tau_{\bf k}} )$, where
\begin{linenomath*}
\begin {equation}
\label {tauk}
\tau_{\bf k}  = \frac {1}  { \mu (1 + \lambda^2 {k}^2)^q} 
\end {equation}
\end{linenomath*}
is the temporal length scale associated with the spatial wavevector ${\bf k}$.

Note that by the spectrum (\eg $b_{\bf k}$) we always mean the {\em modal} spectrum,
\ie the variance  associated with a single basis function
(a single wavevector ${\bf k}$); the modal spectrum
is not to be confused with the variance (or energy) spectrum.

\subsection{Physical-space statistics}

In the stationary regime (\ie after an initial transient period has passed),
the above independence of the spectral random processes $\tilde\xi_{\bf k}(t)$
(see section \ref{sec_order1_spe})
implies that the random field $\xi(t, {\bf s})$ is 
spatio-temporally {\em homogeneous},
\ie invariant under shifts in space and time: 
\begin{linenomath*}
\begin {equation}
\label {hom}
\Ex \xi(t, {\bf s}) \cdot \overline {\xi(t + \Delta t, {\bf s} + \Delta {\bf s})} = B(\Delta t, \Delta {\bf s}),
\end {equation}
\end{linenomath*}
where $\Ex$ is the expectation operator and
\begin{linenomath*}
\begin {equation}
\label {Bts}
B(t, {\bf s}) = \sum_{\bf k}  b_{\bf k}\, C_{\bf k}(t) \,\e^{\i ({\bf k}, {\bf s})}.
\end {equation}
\end{linenomath*}
In particular, the spatial covariance function is
\begin{linenomath*}
\begin {equation}
\label {Bs}
B({\bf s}) = B(t=0, {\bf s})  = 
     \sum_{\bf k}  b_{\bf k} \,\e^{\i ({\bf k}, {\bf s})},
\end {equation}
\end{linenomath*}
where it is seen that the spatial spectrum $b_{\bf k}$ is the Fourier transform of the 
spatial covariance function  $B({\bf s})$.

The temporal covariance function is
\begin{linenomath*}
\begin {equation}
\label {B_t}
B({t}) = B(t, {\bf s=0})  = 
     \sum_{\bf k}  b_{\bf k} \,C_{\bf k}(t).
\end {equation}
\end{linenomath*}
Finally, the variance is 
\begin{linenomath*}
\begin {equation}
\label {Vxi}
\Var\xi = B(t=0, {\bf s=0})  =  \sum_{\bf k}  b_{\bf k}.
\end {equation}
\end{linenomath*}
%

\subsection{``Proportionality of scales'' requires that  $q=\frac12$}
\label{sec_PSproperty}

The more precise formulation of the ``proportionality of scales'' requirement \ref{req_SPG_PS} 
states that for large $k$,
the temporal length scale $\tau_{\bf k}$ should be inversely proportional to $k$:
\begin{linenomath*}
\begin {equation}
\label {cond_PS}
\tau_{\bf k} \sim \frac{1}{k} \quad \mbox{as} \quad k \to \infty.
\end {equation}
\end{linenomath*}
From Eq.(\ref{tauk}), this condition entails, importantly, that
\begin{linenomath*}
\begin {equation}
\label {cond_PS2}
q=\frac12.
\end {equation}
\end{linenomath*}
Below, we show that the choice $q=\frac12$ causes the generated spatio-temporal random fields
to possess, besides the ``proportionality of scales'', many other nice properties (sections 
\ref{sec_isotr} and \ref{sec_covs}).

\subsection{The spatial operator of order $q=\frac12$}
\label{sec_Aq12}

The model's spatial operator $A$ becomes (see Eq.(\ref{ALapl2}))
\begin{linenomath*}
\begin {equation}
\label {ALapl12}
A = \mu ({1 - \lambda^2\Delta})^\frac12 \equiv \mu \sqrt{ 1 - \lambda^2\Delta }. 
\end {equation}
\end{linenomath*}
This is  a pseudo-differential operator \citep[e.g.][]{Shubin} with the {\em symbol} 
\begin{linenomath*}
\begin {equation}
\label {symb}
a({k}) =  \mu \sqrt{1 + \lambda^2{k}^2}, 
\end {equation}
\end{linenomath*}
so that the action of $A$
on the test function $\varphi({\bf s})$ is defined as follows. 
First, we Fourier transform $\varphi({\bf s})$ getting $\{ \tilde\varphi_{\bf k} \}$.
Then,  $\forall {\bf k} \in \Z^d$, we multiply $\tilde\varphi_{\bf k}$ by the symbol 
$a({k})$.
Finally, we perform the backward Fourier transform of
$\{ a({k}) \tilde\varphi_{k} \}$
retrieving the result, the function $(A\varphi)({\bf s})$.

So, the action of the above fractional negated and shifted Laplacian on test functions in spectral space 
is well defined.
Importantly, in physical space, the pseudo-differential operator $A$  can  be 
approximated by a discrete-in-space linear operator which is represented by a {\em very sparse matrix},
see Appendix \ref{app_symbol}.
So, in both spectral space and physical space, the resulting operator $A$ with the
fractional degree $q=\frac12$ is numerically tractable.

\subsection{The first-order model cannot satisfy the SPG requirements}
\label{sec_impose}

Let us compute $\Var\xi$ using Eqs.\ref{bk}  and \ref{Vxi}.
Since $b_{\bf k}$  is a smooth function of the wavevector ${\bf k}$, we may approximate the sum in Eq.\ref{Vxi}
with the integral (where $b({\bf k})=b_{\bf k}$ for integer wavenumbers), getting
\begin {equation}
\label {fvd}
\Var\xi \propto  \int _{\R^d}   \frac {1}  {\sqrt{1 + \lambda^2 {k}^2}} \,\d {\bf k} \propto 
       \int_{\R} \frac {k^{d-1}}  {\sqrt{1 + \lambda^2 {k}^2}} \,\d k.
\end {equation}
To  check the  convergence of the latter integral in Eq.(\ref{fvd}), 
we examine the $k \to \infty$ limit. For large $k$, the integrand is, obviously,
proportional to $k^{d-2}$.
As we know, the integral of this kind converges if the integrand decays faster than ${k^{-1-\epsilon}}$
with some $\epsilon >0$.
This implies that the integral in Eq.(\ref{fvd}) diverges for all $d \ge 1$.
In other words, 
the spectrum Eq.(\ref{bk}) decays too slowly for $\Var\xi$ to be finite. 

So, the SPG model Eq.(\ref{bpg1}) cannot simultaneously satisfy the proportional-scales requirement  \ref{req_SPG_PS} 
(which leads to $q=\frac12$) and the finite-variance  requirement  \ref{req_SPG_cont}.
Consequently, the SPG model is to be somehow changed.
The solution is to increase the temporal order of the model.

\section {Higher-order in time model}
\label{sec_phy}

\subsection{Formulation}
\label{sec_motiv}

The   SPG model of higher temporal order  is
\begin{linenomath*}
\begin {equation}
\label {bpg_p}
\left( \frac {\partial }{\partial t} +  \mu \sqrt{1 - \lambda^2\Delta} \right)^p  \xi(t,{\bf s}) =  
\alpha(t,{\bf s}),
\end {equation}
\end{linenomath*}
where $p$ is the temporal order of the modified SPG model (a positive integer).
In spectral space, the  model  reads (cf. section \ref{sec_order1_spe})
\begin{linenomath*}
\begin {equation}
\label {bpg_ps}
\left(\frac{\d} {\d t} +  \mu  \sqrt{1 + \lambda^2 k^2}\right)^p  \tilde\xi_{\bf k}(t) =
       \sigma \, \Omega_{\bf k}(t).
\end {equation}
\end{linenomath*}
In this section, we explore the steady-state statistics of $\xi(t,{\bf s})$ 
and find out which values of the temporal order $p$ solve the above infinite variance problem.

\subsection{Stationary spectral-space statistics}

For each ${\bf k}$, Eq.(\ref{bpg_ps}) is a $p$th-order  in time OSDE. 
Using Table \ref{Tab_osde} in Appendix \ref{app_OSDE},
we can write down
the stationary variance $b_{\bf k}$  and the temporal correlation function $C_{\bf k}(t)$
of the solution to Eq.(\ref{bpg_ps}), the process  $\tilde\xi_{\bf k}(t)$:
 \begin{linenomath*}
\begin {equation}
\label {bk12_p}
 b_{\bf k}  \propto \frac {\sigma^2}  { \mu^{2p-1} (1 + \lambda^2 {k}^2)^{p-\frac12} } 
\end {equation}
\end{linenomath*}
(where the sign $\propto$ means proportional to)
and
 \begin{linenomath*}
\begin {equation}
\label {Ck_p}
 C_{\bf k}(t)  = \left(1 + \frac{|t|} {\tau_{\bf k}}  + r_2  \frac{|t|^2} {\tau_{\bf k}^2} + \dots + 
               r_{p-1}  \frac{|t|^{p-1}} {\tau_{\bf k}^{p-1}} \right) 
               \,\e^{- \frac{|\mathnormal{t}|}{\tau_{\bf k}}  }.
\end {equation}
\end{linenomath*}
Here $r_2,\dots, r_{p-1}$ are real numbers (given for $p=1,2,3$ in Table \ref{Tab_osde},
see Appendix \ref{app_OSDE})
and $\tau_{\bf k}$ are still defined by Eq.(\ref{tauk}).
Specifically, for the temporal order  $p=3$, we have
 \begin{linenomath*}
\begin {equation}
\label {bk_p3}
 b_{\bf k}|_{p=3} = \frac {3\sigma^2}  {16 \mu^5 (1 + \lambda^2 {k}^2)^\frac{5}{2} }
\end {equation}
\end{linenomath*}
and
 \begin{linenomath*}
\begin {equation}
\label {Ck_p3}
 C_{\bf k}(t)|_{p=3}  = \left(1 + \frac{|t|} {\tau_{\bf k}}   + \frac13 \,\frac{|t|^{2}} {\tau_{\bf k}^{2}} \right) \,\e^{- \frac{|\mathnormal{t}|}{\tau_{\bf k}}  }.
\end {equation}
\end{linenomath*}
As  Eq.(\ref{tauk}) is unchanged in the higher order model,
the  ``proportionality of scales'' condition Eq.(\ref{cond_PS}) is still satisfied.
In order to achieve the desired dependency of $\tau_{\bf k}$ not only on $k$ (which we already have from Eq.(\ref{tauk})),
but also on $\lambda$ (the greater is $\lambda$  the greater should be $\tau_{\bf k}$), 
we parameterize $\mu$ as
 \begin{linenomath*}
\begin {equation}
\label {mu}
 \mu   =  \frac{U} {\lambda},
\end {equation}
\end{linenomath*}
where $U>0$ is the velocity-dimensioned tuning parameter.
Note that $\lambda$ affects both the spatial length scale of $\xi$
(due to Eq.(\ref{bk12_p})) and the temporal length scale (thanks to Eq.(\ref{tauk})).
In contrast, $U$ affects only the temporal length scale.

\subsection{Finite-variance criterion}
\label{sec_finvar_p}

Substituting $b_k$ from Eq.(\ref{bk12_p}) into Eq.(\ref{Vxi}),
approximating the sum over the wavevectors by the integral,
and exploiting the isotropy of the integrand yields
\begin{linenomath*}
\begin {equation}
\label {vardp}
\Var\xi \approx \constant \cdot \int_0^\infty \frac{\sigma^2}  {  (1 + \lambda^2 {k}^2)^{p-\frac12} }  \, k^{d-1} \,\d k, 
\end {equation}
\end{linenomath*}
so that we have $\Var\xi <\infty$ (requirement \ref{req_SPG_cont}) whenever
\begin{linenomath*}
\begin {equation}
\label {critdp}
p > \frac{d+1}{2}. 
\end {equation}
\end{linenomath*}
%

\subsection{Isotropy}
\label{sec_isotr}

In this section, we show that, remarkably, $q=\frac12$ is the unique spatial order for which the field $\xi(t, {\bf s})$ 
appears to be isotropic in space-time. In particular, the shape of the correlation function
is the same in any spatial or temporal or any other direction in the spatio-temporal domain
$\T^\mathnormal{d} \times \R$.

\subsubsection{Spatial isotropy}
\label{sec_spat_isotr}

We note that the spatial isotropy of the random field $\xi$ 
is the invariance of its covariance function $B({\bf s})$ under rotations.
If we were in $\R^d$ rather than on $\T^\mathnormal{d}$, 
isotropy of $B({\bf s})=B(s)$, where $s = |{\bf s}|$ is the spatial distance, would be equivalent 
to isotropy of its Fourier transform (spectrum) $ b({\bf k})$, so that the latter would be
dependent only  $k=|{\bf k}|$.
On the torus, spectra are discrete, i.e. $m,n,l$ take only {\em integer} values,
so, strictly speaking,  $b_{\bf k}$ cannot be isotropic there.
To avoid this technical difficulty, we resort (for the theoretical analysis only)
to the device used in sections \ref{sec_impose} and
\ref{sec_finvar_p}, the approximation of a sum over
the wavevectors by an integral.

Specifically, we assume that  $b_{\bf k}$ is smooth enough
(which is tantamount to the assumption that $B({\bf s})$ decays on length scales much smaller than
the domain's extents) for the validity of the approximation
\begin{linenomath*}
\begin {equation}
\label {Bs_int}
B({\bf s})  = \sum_{\bf k \in \Z^d}  b_{\bf k} \,\e^{\i ({\bf k}, {\bf s})} \approx
                  \int_{\R^d} b({\bf k}) \,\e^{\i ({\bf k}, {\bf s})} \d {\bf k},
\end {equation}
\end{linenomath*}
where $b({\bf k})$ is a smooth function of the real vector argument ${\bf k} \in \R^d$ such that
$\forall \bf k \in \Z^d, b({\bf k})=b_{\bf k}$.
The integral in  Eq.(\ref{Bs_int}) with the isotropic $b({\bf k})$, see  Eq.(\ref{bk12_p}),
can be easily shown to be invariant under rotations of ${\bf s}$.
This implies  that $B({\bf s})$ and so the random field $\xi$ are 
indeed approximately spatially isotropic.

In the theoretical analysis in this section, we will rely on the approximation Eq.(\ref{Bs_int})
and thus assume that the ``spectral grid'' is dense enough for the spatial spectra to be treated
as continuous ones.

\subsubsection{Isotropy in space-time}
\label{sec_spatemp_spec}

Consider the OSDE Eq.(\ref{bpg_ps}) in the stationary regime. 
Following \citet[][section 8]{Yaglom}, the stationary random process can be spectrally
represented as the stochastic integral
\begin{linenomath*}
\begin {equation}
\label {timspe}
\tilde\xi_{\bf k}(t) = \int_{\R} \e^{\i \omega \mathnormal{t}} \, Z_{\bf k}(\d \omega),
\end {equation}
\end{linenomath*}
where $\omega$ is the angular frequency (temporal wavenumber) and $Z$ is the 
orthogonal stochastic measure such that 
\begin{linenomath*}
\begin {equation}
\label {b_k_omega}
\Ex |Z_{\bf k}(\d \omega)|^2 = b_{\bf k}(\omega) \, \d \omega,
\end {equation}
\end{linenomath*}
where $b_{\bf k}(\omega)$ is the spectral density of the process $\tilde\xi_{\bf k}(t)$ 
(\ie the Fourier transform of its covariance function $b_{\bf k}  C_{\bf k}(t)$,
see Eq.(\ref{bt})) and, 
at the same time, the spatio-temporal spectrum of the field $\xi$.
In the spectral expansion of the driving white noise $\Omega_{\bf k}(t)$ (see Eq.(\ref{bpg_ps})),
\begin{linenomath*}
\begin {equation}
\label {wn_spe}
 \Omega_{\bf k}(t) =  \int_{\R} \e^{\i \omega \mathnormal{t}} \, Z_{\Omega_{\bf k}}(\d \omega),
\end {equation}
\end{linenomath*}
we have $\Ex |Z_{\Omega_{\bf k}}(\d \omega)|^2 =\constant \cdot \d\omega$ because the white noise  
has constant spectral density.
Next, we substitute Eqs.(\ref{timspe}) and (\ref{wn_spe}) into Eq.(\ref{bpg_ps}), getting
\begin{linenomath*}
\begin {equation}
\label {ZZ}
 (\i\omega + \mu  \sqrt{1 + \lambda^2 k^2})^p Z_{\bf k}(\d \omega) =  Z_{\Omega_{\bf k}}(\d \omega).
\end {equation}
\end{linenomath*}
In this equation, taking expectation of the squared modulus of both sides,
recalling that $\mu=U/\lambda$,
 and introducing the scaled angular frequency $\omega' = \omega /U$, we finally obtain
\begin{linenomath*}
\begin {equation}
\label {b_k_omega_p}
b_{\bf k}(\omega') \equiv b_{\bf K}  \propto \frac{1}{ (\lambda^{-2} + (\omega')^2 +  k^2)^p } = 
                                     \frac{1}{ (\lambda^{-2} + {\bf K}^2)^p },
\end {equation}
\end{linenomath*}
where 
\begin{linenomath*}
\begin {equation}
\label {Kk}
{\bf K} = (\ \frac{\omega}{U}, {\bf k}) \equiv (\frac{\omega}{U}, m, n, l )
\end {equation}
\end{linenomath*}
is the spatio-temporal wavevector.

From Eq.(\ref{b_k_omega_p}), one can see that with the scaled frequency (note that the change 
$\omega \to \omega / U$ corresponds to the change of the time coordinate
$t \to t \cdot U$), the spatio-temporal spectrum $b_{\bf k}(\omega') \equiv b_{\bf K}$ becomes
{\em isotropic in space-time}. This implies that the correlation function of $\xi$ is isotropic 
in space-time as well (with the scaled time coordinate).
Note that this remarkable property can be achieved only with the spatial order $q=\frac12$.


It is worth noting that
the functional form of the spatio-temporal spectrum Eq.(\ref{b_k_omega_p})
 together with the constraint Eq.(\ref{critdp}) imply that the conditions of Theorem 3.4.3 in 
\citet[][]{Adler} are satisfied, so that spatio-temporal sample paths of the random field $\xi$ are almost surely continuous,
as we demanded in section \ref{sec_req}, see requirement \ref{req_SPG_cont}.

\subsection {Spatio-temporal covariances: the Mat\'ern class}
\label{sec_covs}

The spatio-temporal  field satisfying the $p$-th order SPG model Eq.(\ref{bpg_p}) has the spatio-temporal 
correlation function belonging to the so-called {\em Mat\'ern} class of 
covariance functions \citep[e.g.][]{SteinBook,Guttorp}. To see this, we denote
 \begin{linenomath*}
\begin {equation}
\label {nupd}
\nu = p - \frac{d+1}{2} >0,
\end {equation}
\end{linenomath*}
where the positivity follows from Eq.(\ref{critdp}).
Then Eq.(\ref{b_k_omega_p})  rewrites as
\begin{linenomath*}
\begin {equation}
\label {bk_p_nu}
b_{\bf K} \propto \frac {1}  { (\lambda^{-2} +  {\bf K}^2)^{\nu+\frac{d+1}{2}} }.
\end {equation}
\end{linenomath*}
Note that here $d+1$ is the dimensionality of space-time.
Equation (\ref{bk_p_nu}) indeed presents the  spectrum of the 
Mat\'ern family of correlation functions, see \eg Eq.(32) in \citet{SteinBook}.
The respective isotropic correlation function is given by the equation that precedes Eq.(32)
in \citet{SteinBook} or by Eq.(1) in \citet[][]{Guttorp}:
\begin{linenomath*}
\begin {equation}
\label {matern}
B(r)  \propto ({r}/{\lambda})^\nu K_\nu ({r}/{\lambda}),
\end {equation}
\end{linenomath*}
where $r=\sqrt{s^2 + (Ut)^2}$ is the distance 
(in our case, the Euclidean distance in space-time with the coordinates $(x,y,z,Ut)$)
and $K_\nu$ is the MacDonald function (the modified Bessel function of the second kind).

The Mat\'ern family is often recommended
for use in spatial analysis due to its notable flexibility with only two
free parameters: $\nu$ and $\lambda$, see \eg \citet{SteinBook} and \citet{Guttorp}.
Specifically, $\lambda$ controls the  length scale, whereas
$\nu>0$ determines the degree of smoothness: the higher $\nu$, the 
smoother the field. Note that the smoothness is
understood as the number of the mean-square derivatives of the random field in question.
The degree of smoothness
depends on the behavior of the correlation function at small distances
and manifests itself in field's realizations as the amount of small-scale noise
(for illustration see Appendix \ref{app_smoo}).

Table \ref{Tab_Matern} lists the resulting correlation functions (in any direction
in space-time) for 
several combinations of $d$ and $p$ \citep[see][for details]{Guttorp}.

\begin{table}
\caption{\bf Spatio-temporal correlation functions $B(r)$ for some plausible combinations of
the dimensionality  $d$ and the temporal order $p$}

\begin{center}
\begin{tabular}{|c|c||c||c|}

\hline
  $d$ &  $p$  &  $\nu=p - \frac{d+1}{2}$    & $B(r)$\\
      \hline
  2  &  2  &    $\frac12$ &   $\e^{-\frac{\mathnormal{r}}{\lambda}}$   \\
      \hline
  2  &  3  &    $\frac32$ &   $(1 + \frac{r}{\lambda})\,\e^{-\frac{\mathnormal{r}}{\lambda}}$   \\ 
      \hline
  2  &  4  &    $\frac52$ &   $(1 + \frac{r}{\lambda} + \frac13 \left(\frac{r}{\lambda}\right)^2)\,\e^{-\frac{\mathnormal{r}}{\lambda}}$   \\ 
        \hline
  3  &  3  &    $1$ &   $\frac{r}{\lambda} K_1 (\frac{r}{\lambda})$   \\
\hline
\end{tabular}
\end{center}
\label{Tab_Matern}
\end{table}

With the fixed $d$, the larger $p$ corresponds, according to Eq.(\ref{nupd}),
to the larger $\nu$ and so to the smoother in space and time field $\xi$.
This can be used to change the degree of smoothness of the generated field by changing
the temporal order of the SPG model.

From the constraint Eq.(\ref{critdp}), the minimal temporal order $p$ that can be used in both 2D and 3D is 
equal to 3. {\em This value ${p=3}$ will be used by default in what follows and in the current SPG computer program}.

\subsection{Spatio-temporal correlation functions: illustrations}
\label{sec_spatim_illu}

Here we show spatial, temporal, and spatio-temporal correlation functions
computed using Eq.(\ref{matern}).
To make the plots more accessible, it is arbitrarily assumed  that the
extent of the standardized spatial domain (the torus) in each dimension is 3000 km, so that
the distance is measured in kilometers.
The  default SPG  setup parameters are 
$\lambda=125$ km and $U=20$ m/s.

\subsubsection{Spatial correlation functions}
\label{sec_spacrf}

\begin{figure}
\begin{center}
   { \scalebox{0.3}{ \includegraphics{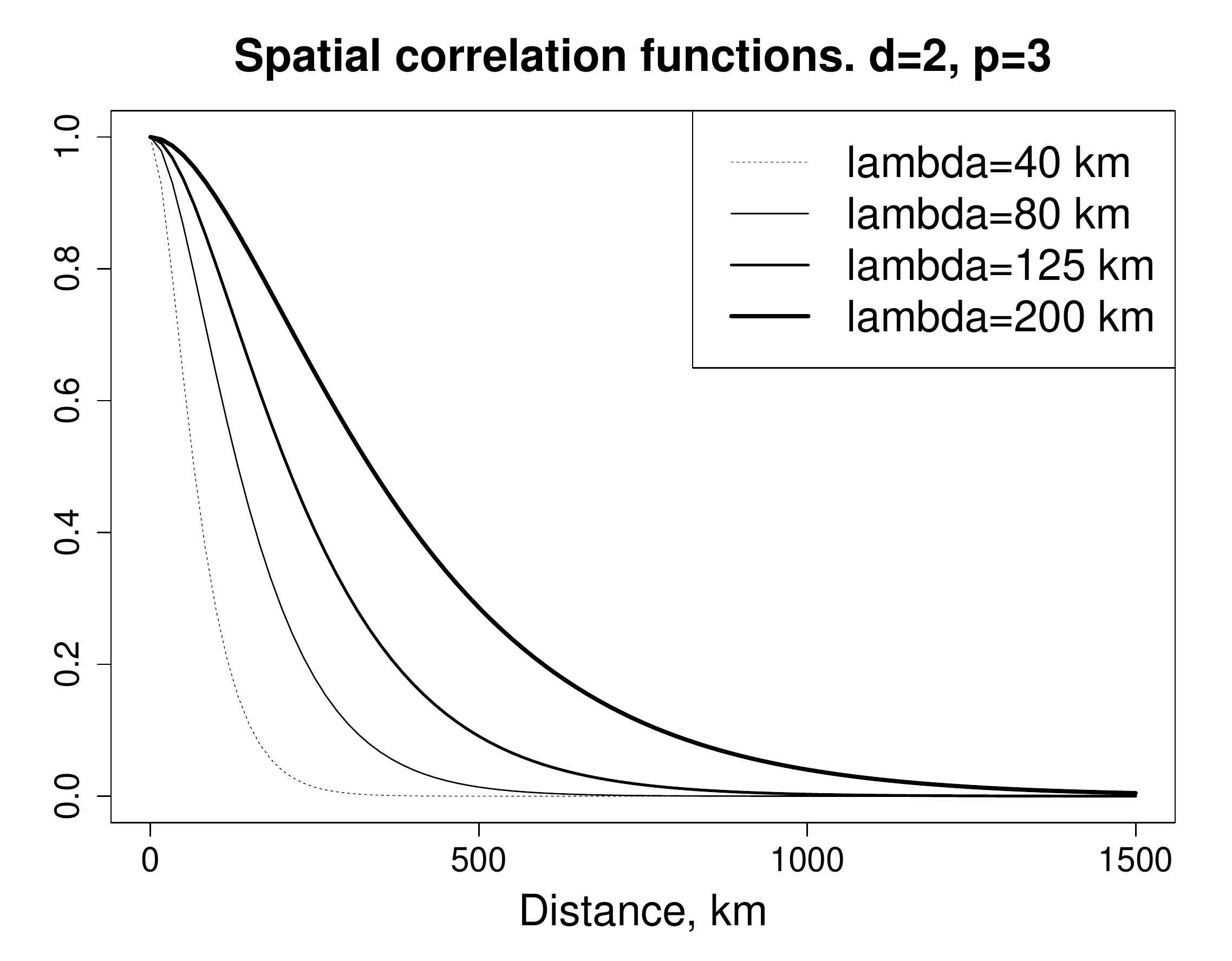}}
     \scalebox{0.3}{ \includegraphics{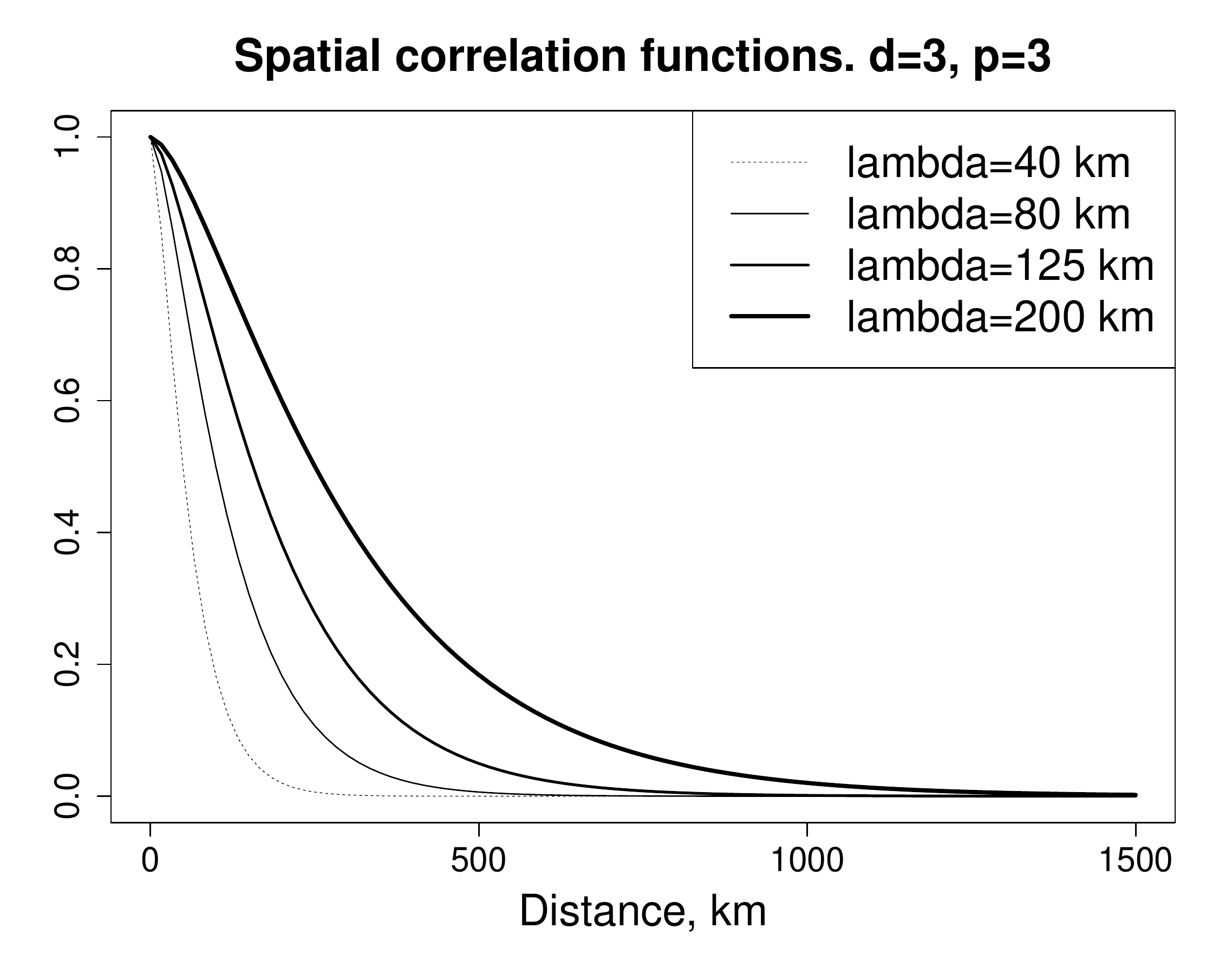}}
       }
\end{center}
  \caption{Spatial correlation functions for $p=3$ in 2D (the left panel) and 3D (the right panel)---for
  the four spatial length
  scales indicated in the legend.
  }
\label{Fig_scrf}
\end{figure}

Figure \ref{Fig_scrf} presents the {spatial} correlation functions 
for different length scales in 2D and 3D.
One can notice, first, that the actual length scale is well controlled by the parameter $\lambda$.
Second, it is seen that in 2D (the left panel), where, according to Eq.(\ref{nupd}), $\nu= \frac{3}{2}$, the correlation functions 
are somewhat {\em smoother} at the origin than in 3D (the right panel),  where $\nu=1$.
This is consistent with the above statement that the greater $\nu$ the smoother the field.
But in general, the 2D and 3D spatial correlation functions are quite similar.


%

\subsubsection{Temporal correlation functions}
\label{sec_timcrf}

Equation (\ref{matern}) implies that the spatial and temporal correlations have the same {\em shapes}.
The latter feature is very nice because atmospheric spectra are known to be similar in the spatial and in the temporal domain,
\eg  the well-known ``-5/3'' spectral slope law
is observed both in space and time, see \eg \citet[section 23]{Monin}.
So, the SPG does reproduce this observed in the atmosphere similarity of spatial and temporal correlations.

\begin{figure}
\begin{center}
   { \scalebox{0.3}{ \includegraphics{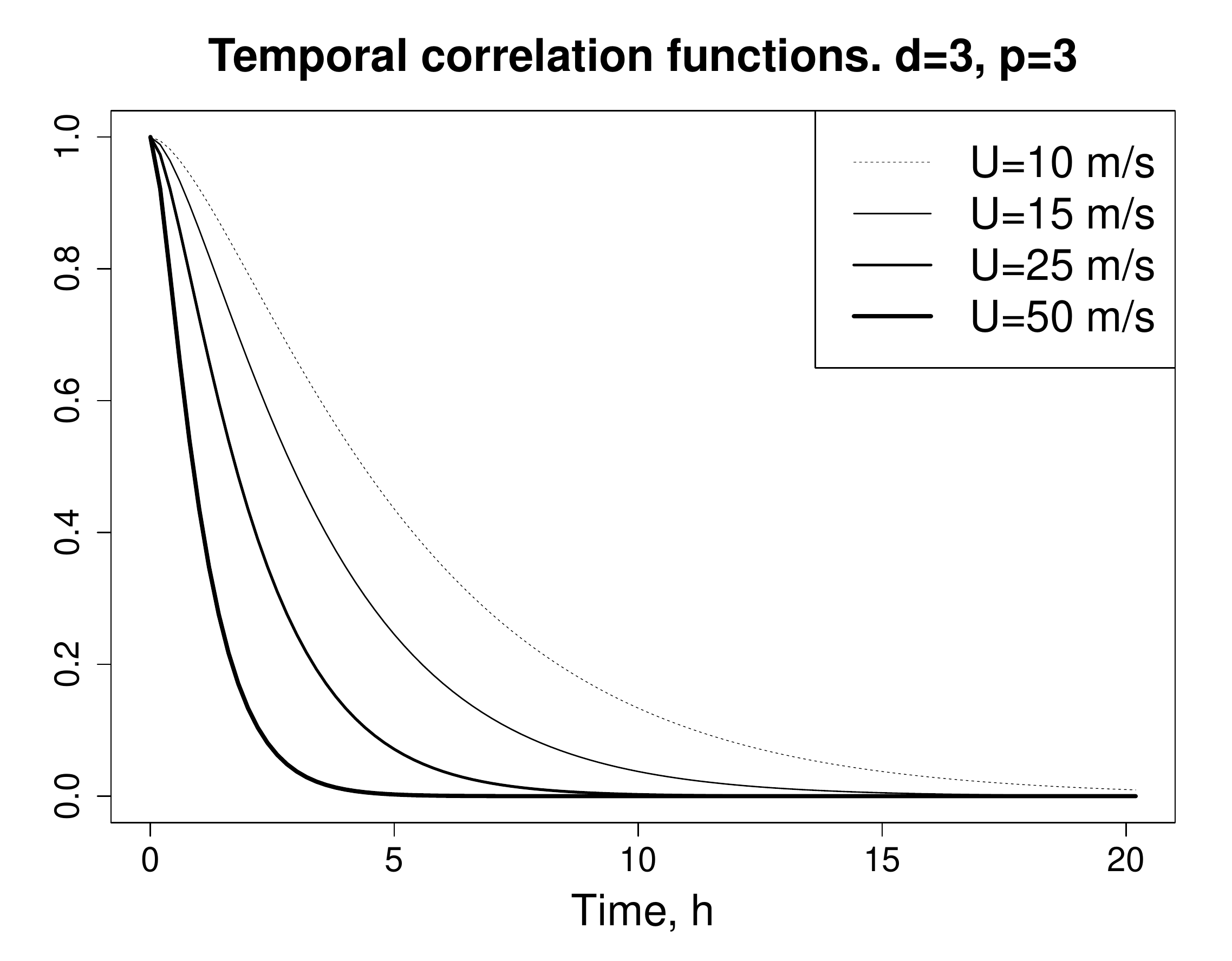}}
       }
\end{center}
  \caption{Temporal correlation functions in 3D for the four values of $U$ indicated in the legend.
  }
\label{Fig_tcrf_U}
\end{figure}

Figure \ref{Fig_tcrf_U} shows the temporal correlation functions 
for different parameters $U$.
Comparing Fig.\ref{Fig_tcrf_U} with  Fig.\ref{Fig_scrf}(right), one can observe
that the spatial and temporal correlations indeed have the same shape.


\subsubsection{Spatio-temporal correlations}
\label{sec_spatim}

\begin{figure}
\begin{center}
   { \scalebox{0.3}{ \includegraphics{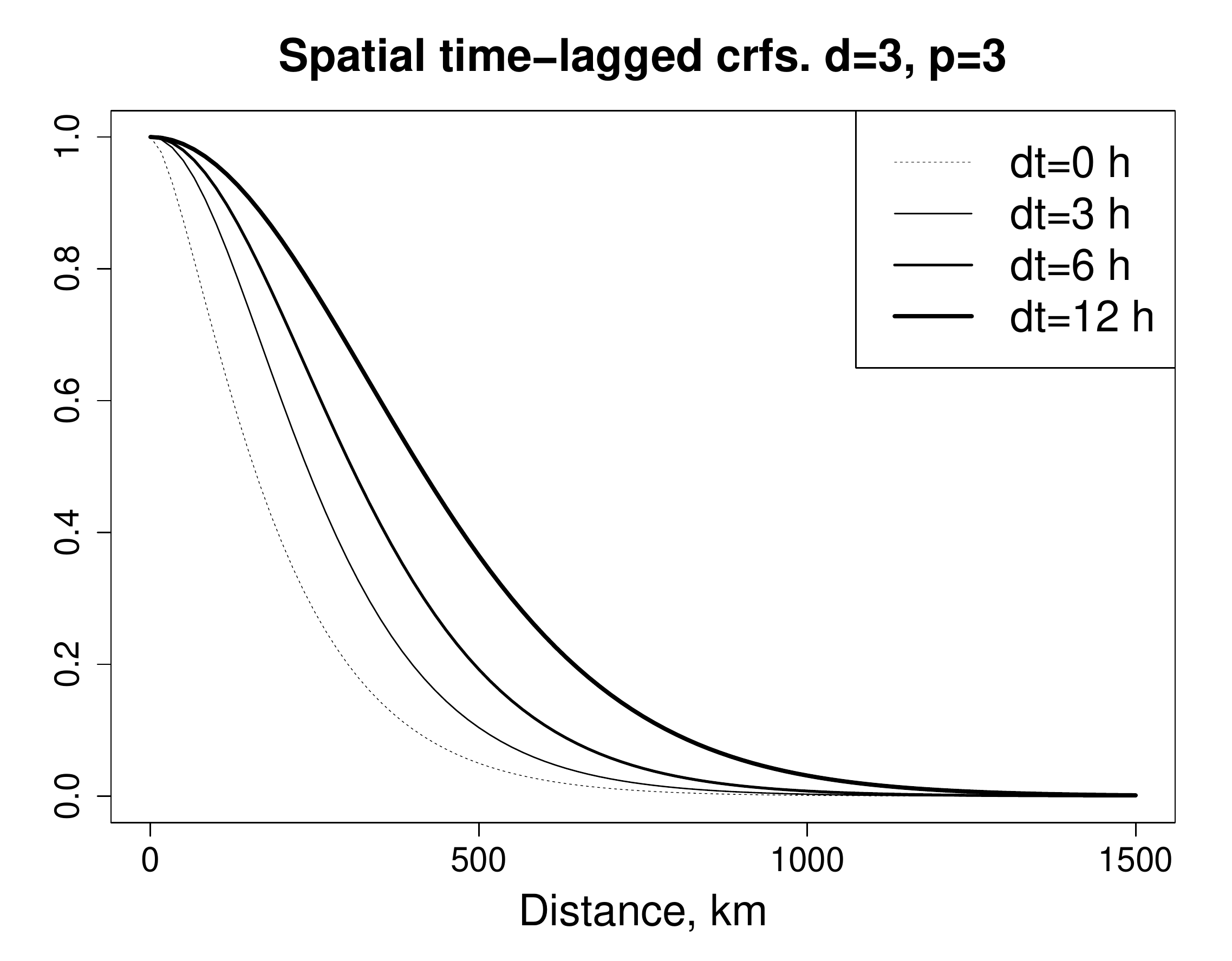}}
       }
\end{center}
  \caption{Spatial correlation functions in 3D for the four time lags indicated in the legend.
  }
\label{Fig_spatim1}
\end{figure}

\begin{figure}
\begin{center}
   { \scalebox{0.3}{ \includegraphics{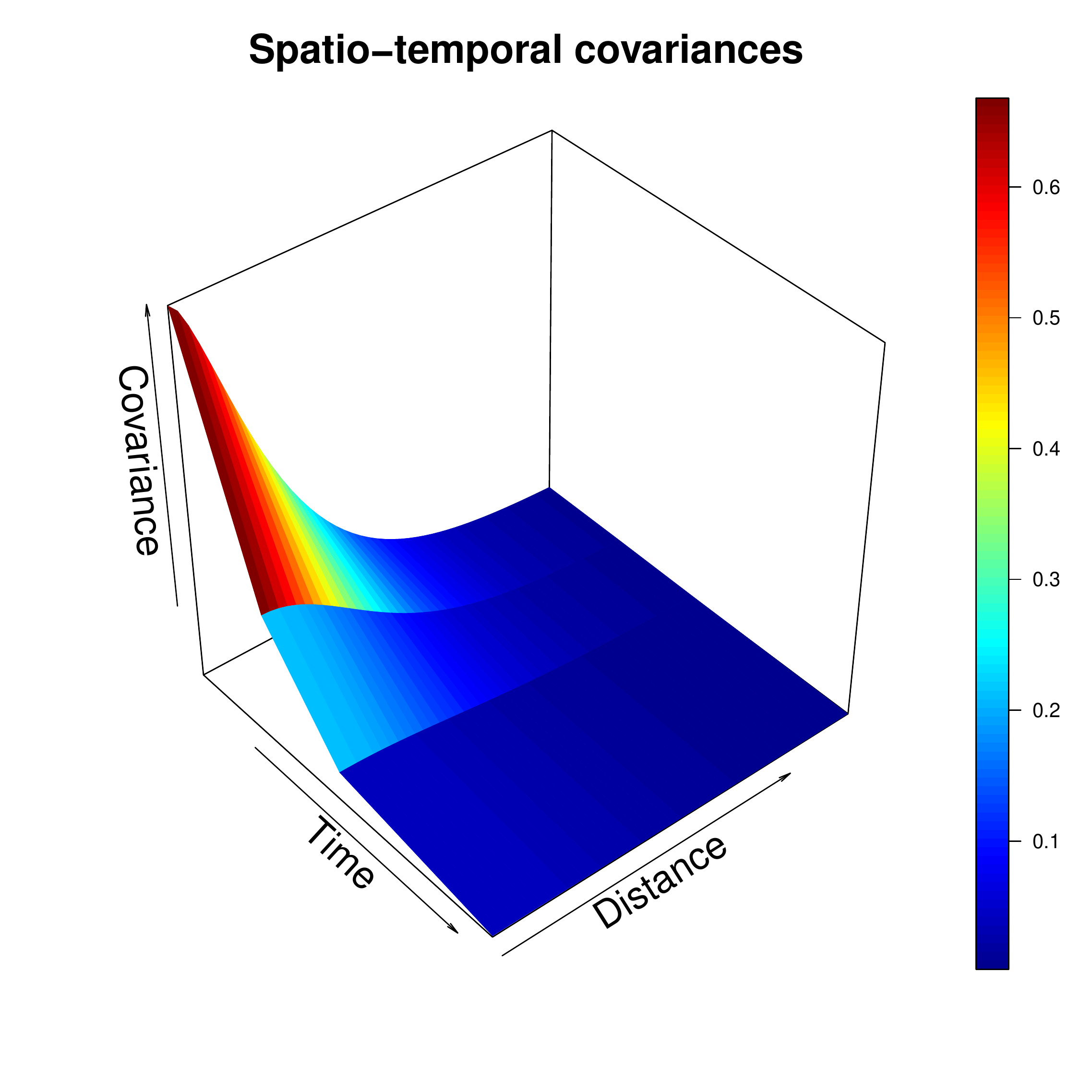}}
       }
\end{center}
  \caption{Spatio-temporal SPG covariances.
  }
\label{Fig_spatim2}
\end{figure}

Figure \ref{Fig_spatim1} presents the spatial correlation functions for different
time lags. Figure \ref{Fig_spatim2}  displays the spatio-temporal correlation function.
In both Fig.\ref{Fig_spatim1} and Fig.\ref{Fig_spatim2},
another manifestation of the spatio-temporal ``proportionality of scales'' is seen:
the larger the time lag, the broader the spatial correlations.
Note that this is consistent with the behavior of the spatio-temporal covariances
  found by \citet[Fig.8]{CressieHuang} in real-world wind speed data.

%


%

\subsection{Introducing anisotropy in the vertical plane}
\label{sec_anis}

We have formulated the SPG model under the 3D isotropy assumption.
This implies that the ratio of the horizontal length scale to the horizontal domain size 
is the same as the ratio of the vertical length scale to the vertical domain size.
This may be reasonable but, obviously, the independent specification of the horizontal
and vertical length scales would be much more flexible.
To get this capability, we can employ two equivalent modifications to the SPG model.
One approach is to change the radius of the ``vertical circle'' in the torus
from $1$ to the  $\delta^{-1}$, where $\delta$ is a positive parameter.
Another approach is to replace the Laplacian 
$\Delta = \frac{\partial^2}{\partial x^2} +  \frac{\partial^2}{\partial y^2} +  \frac{\partial^2}{\partial z^2}$
by its anisotropic version
$\Delta' = \frac{\partial^2}{\partial x^2} +  \frac{\partial^2}{\partial y^2} +  \delta^2\frac{\partial^2}{\partial z^2}$.
With both approaches, the vertical length scale increases by the factor of $\delta$.

\subsection{Preserving isotropy in the horizontal plane for non-square domains}
\label{sec_horz_isotr}

If the size of the domain in physical space in the $x$ direction, $D_x$, differs from the domain size
 in the $y$ direction, $D_y$, then mapping from 
a square SPG domain to the rectangular physical domain would result in an elliptic (also called geometric)
 anisotropy in the horizontal plane.
This undesirable feature can be avoided by replacing $\Delta'$ defined in  section \ref{sec_anis}
with 
\begin{linenomath*}
\begin {equation}
\label {Lapl_anis}
\Delta_* = \frac{\partial^2}{\partial x^2} +  \gamma^2 \frac{\partial^2}{\partial y^2} +  
             \delta^2\frac{\partial^2}{\partial z^2},
\end {equation}
\end{linenomath*}
where $\gamma={D_x} /{D_y}$.
The only change in all  the above  spectral equations is that the  wavenumbers $n$ and $l$ are 
to be multiplied by $\gamma$ and $\delta$, respectively. The total spatial
wavenumber squared $k^2$ is to be replaced everywhere by its scaled version
\begin{linenomath*}
\begin {equation}
\label {k2_anis}
k^2_*=m^2+ (\gamma n)^2+(\delta l)^2.
\end {equation}
\end{linenomath*}

More technically, in our implementation of the SPG, the number of grid points on the torus,
$n_x^T,n_y^T,n_z^T$,
differs from that on the physical-space domain, $n_x,n_y,n_z$, respectively,  for two reasons.
Firstly, the grid on the torus is defined to have more grid points than in physical space,
let us denote them $n_x^+,n_y^+,n_z^+$.
This is done because on the cube with the  periodic boundary conditions (\ie on the torus) the correlations 
between the opposite sides of the cube are close to 1 due to the periodicity.
In order to avoid these spurious correlations on the physical-space domain, we use only part 
of the grid on the torus 
(specifically, $n_x,n_y,n_z$ contiguous grid points) to map the field to the physical-space domain
grid point to grid point.
Having the user-defined grid sizes in the physical-space domain,   $n_x,n_y,n_z$,
we specify the grid sizes on the torus, $n_x^+,n_y^+,n_z^+$, from the condition
that the resulting correlations between the opposite sides of the domain should be less than 0.2.
Secondly, we somewhat further increase the number of grid points
$n_x^+,n_y^+,n_z^+$  in order for the final grid sizes on the torus, $n_x^T,n_y^T,n_z^T$, 
be multiples of 2,3,5 (as required by the fast Fourier transform software we use).

Then, we find $\gamma$ from the requirement that after the mapping from the torus to the physical-space domain,
the length scales of the horizontal function in the $x$ and $y$ directions be the same.
It is easy to see that this is the case if
%
%
\begin{linenomath*}
\begin {equation}
\label {Lapl_anis2gamma}
\gamma=\frac{n_x^T}{n_y^T}.
\end {equation}
\end{linenomath*}
This  device indeed allows to preserve the horizontal isotropy and
to change the vertical length scale in a broad range (not shown).
We have refrained from  introducing this feature to our basic SPG equations
for the sake of simplicity of presentation.

\subsection{The final formulation of the SPG model}
\label{sec_finSPG}

The temporal order of the SPG model is $\boxed{p=3}$.
The  SPG model is 
\begin {equation}
\label {bpg_3}
\boxed{ \left( \frac {\partial }{\partial t} +  \frac{U}{\lambda} \sqrt{1 - \lambda^2\Delta_*} \right)^3  \xi(t,{\bf s}) =  \alpha(t,{\bf s}) },
\end {equation}
where $\alpha(t,{\bf s})$ is the spatio-temporal  white noise.

In spectral space, each spectral coefficient $\tilde\xi_{\bf k}(t)$ satisfies the equation
\begin {equation}
\label {bpg_3spe}
\boxed{ \left(\frac{\d} {\d t} +  \frac{U}{\lambda}  \sqrt{1 + \lambda^2 k_*^2}\right)^3  \tilde\xi_{\bf k}(t) =   
               \sigma \,\Omega_{\bf k}(t) },
\end {equation}
where $\Omega_{\bf k}(t)$ are mutually independent complex standard white noise processes.
The intensity of the spatio-temporal white noise $\alpha$ is $(2\pi)^{d/2} \,\sigma$.

\section {Time discrete  solver for the third-order in time SPG  model}

In {\em physical space}, our final  evolutionary model Eq.(\ref{bpg_p}) with $p=3$  can be discretized using
the approximation  of the operator $\sqrt{1 - \lambda^2\Delta}$ proposed in
Appendix \ref{app_symbol}. 
The respective physical-space solver looks feasible but we do {\em not} 
examine it in this study.
Below, we present our basic {\em spectral-space} technique.
From this point on, we will consider only the {\em spectral} SPG.

\subsection {The spectral solver}

To numerically integrate the SPG  equations in spectral space, 
we discretize  Eq.(\ref{bpg_ps}) (with $p=3$)  using an implicit scheme. 
The model operator $(\frac{\d}{\d t} + a_{\bf k})^3$,
where $a_{\bf k}=\frac{U}{\lambda}  \sqrt{1 + \lambda^2 k_*^2}$ and
$k_*^2$ is defined in Eq.(\ref{k2_anis}),
is discretized by replacing the 
time derivative $\frac{\d}{\d t}$ with the backward finite difference $\frac{{\cal I - B}}{\Delta t}$,
where $\Delta t$ is the time step,  ${\cal I}$ is the identity operator, and ${\cal B}$ is the backshift operator.
The white noise in the \rhs of Eq.(\ref{bpg_ps}) is discretized using Eq.(\ref{wp1}), where 
$\d W_{\bf k}(t)$ is replaced with
$\Delta W_{\bf k}(t) = W_{\bf k}(t+\Delta t) - W_{\bf k}(t)$, and simulated as a zero-mean  Gaussian random variable with the variance $\Delta t$.
As a result, we obtain the time discrete evolution equation
\begin{linenomath*}
\begin {equation}
\label {osde_num2}
\hat\xi_{\bf k}(i) = \frac{1} {\varkappa^3} 
   \left[ 3 \varkappa^2 \hat\xi_{\bf k}(i-1) -3\varkappa\hat\xi_{\bf k}(i-2)+\hat\xi_{\bf k}(i-3)+ 
     \sigma  {\Delta t}^{\frac52}  \zeta_{{\bf k}t} \right],
\end {equation}
\end{linenomath*}
where $i=0, 1,2, \dots$ denotes the discrete time instance, 
$\varkappa=1+ a_{\bf k} \Delta t$,
and $\zeta_{{\bf k}t} \sim CN(0,1)$ are independent complex standard Gaussian pseudo-random variables 
(for their definition, see Appendix \ref{sec_dscr_wnoise}).
Note that the solution of the time-discrete  Eq.(\ref{osde_num2}) is denoted by the hat, $\hat\xi_{\bf k}(i)$, 
in order to distinguish it from the solution of the time-continuous Eq.(\ref{bpg_ps}), which is denoted by the tilde, $\tilde\xi_{\bf k}(t)$.

It can be shown that the numerical stability of the scheme Eq.(\ref{osde_num2}) 
is guaranteed whenever $\varkappa >1$, which is 
always the case because $a_{\bf k} > 0 $ (see Eq.(\ref{symb})).

Note that the derivation of the numerical scheme for a higher-order (\ie with $p>3$) SPG
model is straightforward: one should just raise the difference operator
$\frac{{\cal I - B}}{\Delta t}$ to a power higher than 3.

\subsection {Correction of spectral variances}
\label{sec_spec_correc}

Because of discretization errors, 
the time discrete scheme Eq.(\ref{osde_num2}) gives rise to the steady-state spectral 
variances $\hat b_{\bf k} = \Var\hat\xi_{\bf k}(i)$ that 
are different from the ``theoretical'' ones, $b_{\bf k}$ (given in Eq.(\ref{bk_p3})).
 The idea is to  correct (multiply by a number) the solution $\hat\xi_{\bf k}(i)$
to Eq.(\ref{osde_num2})
 so that the  steady-state variance of the corrected $\hat\xi_{\bf k}(i)$ be equal to $b_{\bf k}$.
To this end, we derive $\hat b_{\bf k}$ from Eq.(\ref{osde_num2})
(using Eq.(\ref {sch_p3V}) in Appendix \ref{app_OSDE_dscr}) and 
then, knowing the ``theoretical''  $b_{\bf k}$, we  introduce the correction coefficients, 
 $\sqrt{b_{\bf k} / \hat b_{\bf k}}$, to be applied to $\hat\xi_{\bf k}(i)$.
This simple device ensures that for any time step, the spatial spectrum and thus the {\em spatial} covariances are perfect.
 But the {\em temporal} correlations do depend on the time step, this aspect is discussed below in
 section \ref{sec_gamma}.

\subsection {``Warm start'': ensuring stationarity from the beginning of the time integration}

To start the numerical integration of the third-order scheme Eq.(\ref{osde_num2}),
we obviously need three initial conditions. 
If the integration is the continuation of a previous run, then we just take
values of $\hat\xi_{\bf k}(i)$ at the  last three time instances $i$ from that previous run; 
this ensures the continuity of the resulting
trajectory. If we start a new integration, we have to somehow generate values of ${\bf\hat \xi}_{\bf k}(i)$ at $i=1,2,3$,
let us denote them here as the vector 
$\boldsymbol\xi^{ini} = ({\bf\hat \xi}_{\bf k}(1), {\bf\hat \xi}_{\bf k}(2), {\bf\hat \xi}_{\bf k}(3))^\top$.
Simplistic choices like specifying zero initial conditions give rise to a substantial initial transient period,
which distorts the statistics of the generated field in the short time range.

In order to have the steady-state regime right from the beginning  of the time integration
and thus avoid the initial
transient period completely, we simulate $\boldsymbol\xi^{ini}$ as a pseudo-random draw from the 
multivariate Gaussian distribution with  zero mean and the steady-state covariance matrix of  $\hat\xi_{\bf k}(i)$.
In Appendix \ref{app_OSDE_dscr}, we derive the components of this
$3\times 3$ matrix, namely, its diagonal elements 
(all equal to the steady-state variance), see Eq.(\ref{sch_p3V}),
and the  lag-1 and lag-2  covariances, see Eq.(\ref{sch_p3c12}).

\subsection {Computational efficiency}
\label{sec_effic}

In this subsection, we describe two techniques that allow us to significantly decrease
the computational cost of running the spectral SPG.

\subsubsection {Making the time step $\Delta t$ dependent on the spatial wavevector ${\bf k}$}
\label{sec_gamma}

For an ordinary differential equation, the accuracy of a finite-difference scheme 
depends on the time step.
More precisely, it depends on the ratio of the time step $\Delta t$ to the temporal length
scale $\tau$ of the process in question.
For high accuracy,   $\Delta t \ll \tau$ is needed.

In our problem,  $\tau_{\bf k}$ decays with the total scaled wavenumber $k_*$, 
see Eqs.(\ref{tauk}),  (\ref{cond_PS}), and (\ref{k2_anis}).
This implies that for higher $k_*$, smaller time steps are needed.
To maintain the accuracy across the wavenumber spectrum, we choose the time step
to be a portion of the time scale:
\begin{linenomath*}
\begin {equation}
\label {osde_num4}
(\Delta t)_{\bf k} = \beta \tau_{\bf k}.
\end {equation}
\end{linenomath*}
The less $\beta$, the more accurate 
and, at the same time, more time consuming the numerical integration scheme is.

We note that in  atmospheric spectra, small scales have, normally, 
much less variance (energy) than large scales.
But with the constant $\beta$, the computational time would be, on the contrary, spent predominantly on
high wavenumbers (because the latter require a smaller time step  and are much more abundant in 3D or 2D).
So, to save computer time whilst ensuring reasonable overall (\ie for the whole range of wavenumbers) accuracy,
we specify $\beta$ to be wavenumber dependent (growing with the wavenumber) in the following ad-hoc way:
\begin{linenomath*}
\begin {equation}
\label {gamma_k}
 \beta_{\bf k} = \beta_{\rm min} + (\beta_{\rm max}  - \beta_{\rm min}) \left( \frac{k_*} { \max k_*}\right)^2, 
\end {equation}
\end{linenomath*}
where $\beta_{\rm min}$ and $\beta_{\rm max}$ are the tunable parameters.
The choice of the ``optimal''  $\beta_{\rm min}$ and $\beta_{\rm max}$ is discussed just below in section
\ref{sec_coarse}.

\subsubsection {Introduction of a coarse grid in spectral space}
\label{sec_coarse}

Here we propose  another technique to reduce the computational cost of the spectral solver.
The technique exploits the {\em smoothness} of the SPG spectrum $b_{\bf k}$  Eq.(\ref{bk_p3}). 
This smoothness allows us to introduce a {\em  coarse grid in spectral space} and
save a lot of computer time by performing the integration of the time discrete 
spectral OSDEs Eq.(\ref{osde_num2}) only for those wavevectors 
that belong to the coarse grid. The spectral coefficients $\hat\xi_{\bf k}(i)$ are then
 interpolated from the coarse grid to the dense (full) grid in spectral space.

The latter interpolation would introduce correlations between different spectral coefficients $\hat\xi_{\bf k}(i)$,
which would destroy the spatial homogeneity. In order to avoid this, we employ a device used to
generate so-called surrogate time series \citep[][section 2.4.1]{Theiler}.
At each $t$, we multiply the interpolated (\ie  dense-grid) $\hat\xi_{\bf k}(i)$ 
by $\e^{\i \theta_{\bf k}}$,
where 
$\theta_{\bf k}$
are independent {\em random phases}, \ie independent for different ${\bf k}$ random variables
uniformly distributed on the segment $[0,2\pi]$.  
It can be easily seen that this multiplication 
removes any correlation between the spectral coefficients.

Note also that the random phase rotation does not destroy the Gaussianity
because $\hat\xi_{\bf k}(i)$ are complex circularly-symmetric random variables
with uniformly distributed and independent of  $|\hat\xi_{\bf k}(i)|$ arguments (phases)
\citep[e.g.][section A.1.3]{Tse}.

In order to preserve the temporal correlations, we keep the set of $\theta_{\bf k}$
constant during the SPG-model time integration.

The exact spectrum $b_{\bf k}$ after the trilinear (bilinear in 2D) interpolation
of  $\hat\xi_{\bf k}(i)$ from the coarse to the full spectral grid  
is imposed in a way similar to that described
in section \ref{sec_spec_correc} as follows.
At any time instance when we wish to compute the physical space field, 
for each ${\bf k}$ on the full spectral grid, the linearly interpolated value $\check\xi_{\bf k}$ 
is the linear combination of the closest coarse-grid points ${\bf k}_j$:
\begin {equation}
\label {trilin}
\check\xi_{\bf k} = \sum_{j=1}^{2^d} w_j \, \hat\xi_{{\bf k}_j}
\end {equation}
 where $\check{}$ denotes the interpolated value and
  $w_j$ is the interpolation weight (note that the set of the closest coarse-grid points ${\bf k}_j$
 depends, obviously, on ${\bf k}$).
In Eq.(\ref {trilin}), the coarse-grid  variances 
$\Var \hat\xi_{{\bf k}_j}= {b}_{{\bf k}_j}$ are known for all ${\bf k}_j$
from the  spectrum $\{b_{\bf k}\}$, see Eqs.(\ref{bk12_p}) or (\ref{bk_p3}). Therefore, we can find 
$\Var \check\xi_{\bf k}= \sum_j w_j^2 \, b_{{\bf k}_j}$.
Besides, we  know which variance $\check\xi_{\bf k}$ {\em should} have on the fine grid, namely $b_{\bf k}$.
So, we normalize $\check\xi_{\bf k}$ by multiplying it by $\sqrt{b_{\bf k} / (\Var \check\xi_{\bf k})}$, 
thus imposing the exact spatial spectrum for all ${\bf k}$.

Technically, the 3D coarse spectral grid is the direct product of  three 1D grids. 
Any of the  (non-uniform) 1D coarse grids is specified as follows. 
Its $j$th point  is located at the fine-grid wavenumber $n_j$, 
which equals $j$ for $|j| \le n_0$ (where $n_0$ is an integer) and equals the closest integer to 
$n_0(1+\varepsilon)^{|j|-n_0}$ for $|j| > n_0$.
Here, $\varepsilon$ is a tunable small positive number.
In the below numerical experiments, the coarse-grid
parameters were $n_0=20$ and $\varepsilon=0.2$, which resulted in the following 
positive 1D coarse-grid points: 
0
1
2
3
\dots
19
20
24
29
35
42
50
60
72
86
103
124
150
(the 1D grid extent was 300 points and, correspondingly, the maximal wavenumber was 150).
%

\subsubsection {Numerical acceleration: results}
\label{sec_gamma_coarse}

As the two above acceleration techniques guarantee that the {\em spatial} spectrum is always precise,
we  tested how these techniques impacted the {\em temporal} correlations 
and what was the speedup.
We performed  a numerical experiment with the 2D SPG on the grid 
with  $300 \times 300$ points, the  mesh size  $h=7$ km,
and the setup parameters
$\lambda=80$ km,  $U=10$ m/s, and $\delta=\gamma=1$.
The time interval $\Delta t_{\rm FFT}$ between the successive backward Fourier transforms  determines the effective  
resolution of the generated field in time.
To make the temporal resolution consistent with the spatial resolution, we selected $\Delta t_{\rm FFT}$
close to $h/U$, namely, $\Delta t_{\rm FFT}=15$ min.
The computations were performed on a single CPU.

The results  are presented in Table \ref{Tab_timings}.
We compared the non-accelerated scheme with  the constant $\beta=0.1$ and without the sparse spectral grid (the second row)
and the accelerated scheme with $\beta_{\rm min}=0.15$, $\beta_{\rm max}=3$, and with the sparse spectral grid (the third row).
From  column 2, it is seen that
the combined effect of the two numerical acceleration techniques 
 on the cost of the {\em spectral-space} computations  (see column 2)
was dramatic:
the speedup was 66  times as compared to the non-accelerated scheme. 
The contributions of the two above numerical acceleration techniques to the  spectral-space  speedup were comparable in magnitude  (not shown).
Most importantly, this   spectral-space speedup was achieved at the very little cost:
the temporal  length scale $T_{0.5}$ 
(defined as the time shift at which the correlation function first  intersects the 0.5 level)  was distorted
by only 4 \% \wrt  the theoretical model  (column 6).
Note, however, that the cost of the interpolation from the sparse spectral grid (column 3) and of the discrete backward 
Fourier transform (column 4) reduced the total speedup of the 2D SPG to 14 times  (see column 5).

\begin{table}
\caption{\bf CPU times of  2D SPG  computations per 1 h of SPG model time and the relative error in the temporal length scale $T_{0.5}$.}

\begin{center}
\begin{tabular}{|c||c|c|c||c|c|}

\hline
  Accelerators    & CPU spec. & CPU interp.  & CPU FFT & Speedup & Rel.err. $T_{0.5}$\\
      \hline
  NO    &  0.66 &   0      & 0.027 & 1 & 3 \% \\
       \hline
  YES   &  0.010 & 0.012 & 0.027 & 14 &4 \% \\
\hline
\end{tabular}
\end{center}
\small{\em Spec. stands for  spectral-space,  interp. means interpolation from the sparse spectral grid, and FFT is the fast Fourier transform.}

\label{Tab_timings}
\end{table}

In 3D, the  SPG operating on the spatial grid with  $300 \times 300 \times 64$ points, took
40-70 times more  CPU time as compared to the above 2D case,
with the accuracy being similar to that indicated in Table \ref{Tab_timings} (not shown).
The  total speedup was only 8 times due to an increased share of the Fourier transform.

\subsection {Examples of the SPG fields}

\begin{figure}
\begin{center}
   { \scalebox{0.5}{ \includegraphics{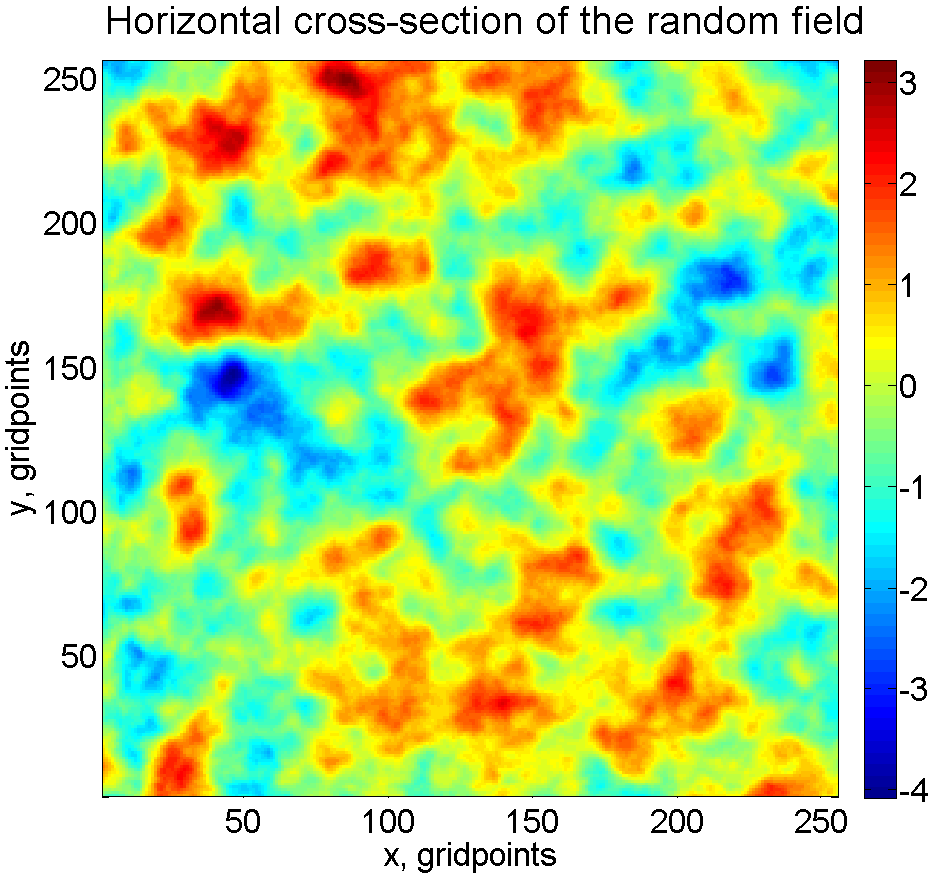}}
       }
\end{center}
  \caption{Horizontal ($x$-$y$) cross-section of an SPG field.}
\label{Fig_xi_xy}
\end{figure}

\begin{figure}
\begin{center}
   { \scalebox{0.5}{ \includegraphics{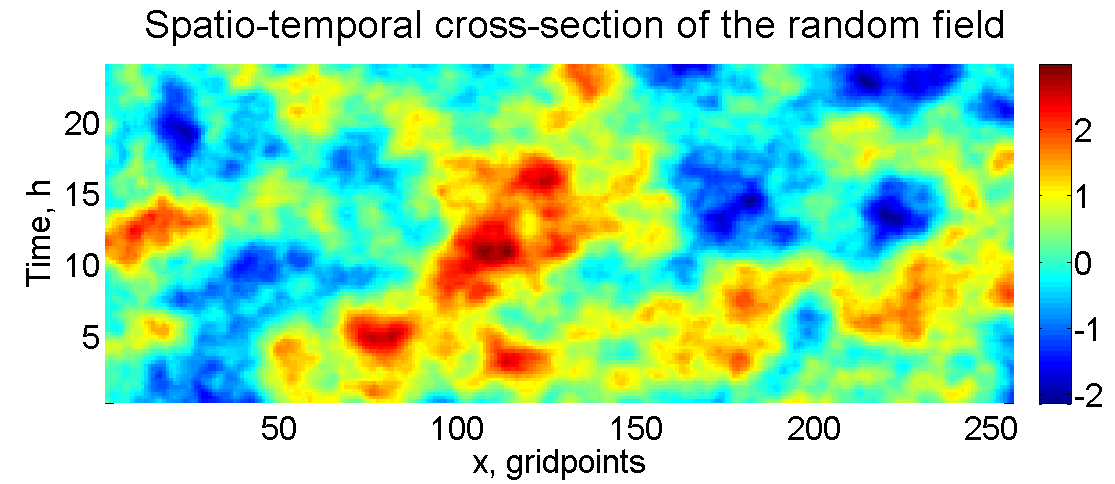}}
       }
\end{center}
  \caption{Spatio-temporal  ($x$-$t$)  cross-section of an SPG field.}
\label{Fig_xi_xt}
\end{figure}

Figure \ref{Fig_xi_xy}  shows a horizontal $x$-$y$ cross-section and 
Fig.\ref{Fig_xi_xt} a spatio-temporal $x$-$t$ cross-section of the  pseudo-random 
field $\xi(t,x,y)$ simulated by the SPG with the setup parameters indicated in section \ref{sec_gamma_coarse}. 
Note that with 300 grid points in each spatial direction, only 256 contiguous grid points are 
shown in the Figs.\ref{Fig_xi_xy} and \ref{Fig_xi_xt} and are intended to be used in a mapping to a physical space domain. 
This is done in the SPG for practical purposes in order to  avoid
 correlations between the opposite sides of the spatial domain, which would be spurious in
 real-world applications.


%

%
%

\section {Discussion}

\subsection {Physical-space or spectral-space  SPG solver?}

In this study, we have investigated both the spectral-space and the physical-space approximations
of the SPG spatio-temporal model. We have found that both approaches can be used to build
a practical SPG scheme. We have selected the spectral-space technique.
Here, we briefly compare both approaches.

Advantages of  the spectral-space technique are the following.
\begin {itemize}
\item
Simplicity of realization. If the SPG model has constant coefficients, then the complicated 
SPG equation decouples into a series of simple OSDEs.

\item
Straightforward accommodation of non-local-in-physical-space spatial  operators.

\end {itemize}

Advantages of  the physical-space approach are:
\begin {itemize}
\item
The relative ease of  introduction of inhomogeneous (non-stationary) and anisotropic 
capabilities to the SPG.

\item
The SPG solver can be implemented in domains with complex boundaries.

\item
Better suitability for an efficient implementation on massively parallel computers.

\end{itemize}

\subsection {Extensions of the  SPG}

The proposed SPG technique can be extended in the future along the following lines.
\begin {itemize}
\item
Development of  a physical-space solver.

\item
Introduction of advection to the SPG model.

\item
Introduction of spatial inhomogeneity/anisotropy and non-stationarity.

\item
Introduction of non-Gaussianity. This can be done either by applying a nonlinear transform to the output SPG fields,
or by introducing a non-Gaussian driving noise \citep[as in][]{Aberg,Wallin}. The former approach is simpler 
but the latter allows for much richer deviations from Gaussianity, including the multi-dimensional aspect.

\item
Going beyond additive and multiplicative perturbations for highly non-Gaussian variables like humidity,
cloud fields, or precipitation.

\item
Simulation of several mutually correlated pseudo-random fields.

\item
Making the temporal order $p$ a user defined variable. As noted above, the larger $p$ the smoother
the generated field.

\end {itemize}

\section {Conclusions}

\begin {itemize}
\item
The proposed Stochastic Pattern Generator (SPG) produces pseudo-random 
spatio-temporal Gaussian fields on 2D and 3D limited area spatial domains
with the tunable variance, horizontal, vertical, and temporal length scales.
\item
The SPG model is defined on a standardized domain in space, specifically, on the unit
2D or 3D torus.
Fields on a limited-area  geophysical domain in question are obtained by mapping from the standardized domain.
\item
The SPG is based on a linear third-order in time stochastic 
model driven by the  white in space and time Gaussian noise.
\item
The spatial operator of the stochastic model is built to ensure that  solutions to the SPG model,
\ie the generated pseudo-random fields, satisfy the  ``proportionality of scales'' property:
large-scale (small-scale) in space field components have large (small) temporal length scales.
\item
Beyond  the ``proportionality of scales'', the generated fields possess a number of other nice properties:
\begin{itemize}
\item
The spatio-temporal realizations are (almost surely) continuous.
\item
With the appropriately scaled time and vertical coordinates, 
the spatio-temporal fields are isotropic in space-time.
\item
The correlation functions in space-time belong to the Mat\'ern class.
\item
The spatial and temporal correlations have the same shapes.
\end{itemize}
\item
The basic SPG solver is spectral-space based.
\item
Two techniques to accelerate the spectral-space computations are proposed and implemented.
The first technique  selects the time step of the spectral-space numerical integration scheme to be 
dependent on the wavenumber, so that the discretization error is smaller for  more energetic larger
spatial scales and is allowed to be larger for less energetic smaller scales.
The second technique introduces a coarse grid in spectral space.
The combined speedup for spectral-space computations 
 from both techniques is as large as 40--60 times.
\item
Potential  applications of the SPG include ensemble prediction and ensemble data assimilation
in meteorology, oceanography, hydrology, and other areas.
The SPG can be used to generate
spatio-temporal perturbations of the model fields (in the additive or multiplicative or other mode)
and of the boundary conditions.
\item
An application of the SPG 
as a source of additive spatio-temporal model error perturbations  
to the meteorological COSMO model \citep{Baldauf}
is described
in  \citep[][]{TsyrulnikovGayfulin_MZ}.
\end {itemize}

\section*{Acknowledgements}
\addcontentsline{toc}{section}{\numberline{}Acknowledgements}%

The SPG has been developed as part of the Priority Project KENDA 
(Kilometre scale Ensemble Data Assimilation) of COSMO.
We have used the discrete fast Fourier package \verb"fft991" developed by C.Temperton at ECMWF in 1978.

\section*{Appendices}
\addcontentsline{toc}{section}{\numberline{}Appendices}%

\appendix

\section{Spatio-temporal structure of the driving 4-D noise}
\label{app_wnoise}

Here, we recall the general definition of the white noise, define the  spatial spectrum 
of the white noise on the $d$-dimensional unit torus, and find its spatial spectral decomposition
in the spatio-temporal case. Then we introduce a colored in space and white in time noise,
and find it spatial spectrum.
Finally, we define the time discrete complex-valued white-noise process.

\subsection{White noise}
\label{app_wnoise_1}

By definition, see \eg \citep[][section 1.1.3]{Rozanov} or \citep[section 3.1.4]{Kuo}, 
the (complex) {\em standard white noise} $\Omega({\bf x})$ defined on a manifold ${\mathbb D}$
is a {\em generalized random field} that acts
on a test function $\varphi({\bf x})$ (where $x \in {\mathbb D}$) as follows:
\begin {equation}
\label {wn1}
(\Omega, \varphi) := \int \varphi({\bf x})\, \Psi(\d {\bf x}),
\end {equation}
where $\Psi$ is the Gaussian orthogonal stochastic measure such that 
for any Borel set $A$,  $\Psi(A)$ is a (complex, in general) Gaussian random variable
with $\Ex \Psi(A) = 0$ and
$\Ex |\Psi(A)|^2 = |A|$,
where $|A|$ is the Lebesgue measure of $A$.

The equivalent definitions of the standard white noise are 
\begin {equation}
\label {wn2}
\Ex|(\Omega, \varphi)|^2 := \int |\varphi({\bf x})|^2 \,\d {\bf x},
\end {equation}
and
\begin {equation}
\label {wn3}
\Ex (\Omega, \varphi) \cdot \overline{(\Omega, \psi)} := 
     \int \varphi({\bf x}) \overline{\psi({\bf x})} \,\d {\bf x},
\end {equation}
where $\psi$ is another test function.
Thus, we have defined of the {\em standard} white noise.
By the general Gaussian white noise, we mean a multiple of the {\em standard} white noise.

\subsection{Spectrum of the  white noise on $\T^\mathnormal{d}$}

The formal Fourier transform of the spatial white noise $\Omega({\bf s})$ 
(where ${\bf s} \in \T^\mathnormal{d}$),
\begin {equation}
\label {wntd}
\tilde\Omega_{\bf k} = \frac{1}{(2\pi)^d} 
   \int_{\T^\mathnormal{d}} \Omega({\bf s}) \,\e^{-\i ({\bf k}, {\bf s})} \,\d {\bf s},
\end {equation}
can be rigorously justified as the action of the white noise $\Omega$ on the test
function 
\begin {equation}
\label {wntd_chi}
\chi({\bf s}) := \frac{1}{(2\pi)^d} \,\e^{-\i ({\bf k}, {\bf s})}.
\end {equation}
Then, the spatial spectrum of $\Omega({\bf s})$ is
\begin {equation}
\label {wntd_b}
b_{\bf k} := \Ex |\tilde\Omega_{\bf k}|^2 \equiv \Ex |(\Omega, \chi)|^2 =
              \int_{\T^\mathnormal{d}} |\chi({\bf s})|^2 \,\d {\bf s} =  \frac{1}{(2\pi)^d}.
\end {equation}
Here, the third equality is due to Eq.(\ref{wn2}).
We stress that it is the {\em modal} spectrum that is constant for the white noise
(not the variance spectrum).

\subsection{Space-integrated spatio-temporal white noise on $\T^\mathnormal{d} \times \R$}

Let us consider the spatio-temporal white noise
$\Omega=\Omega(t, {\bf s})$, where $t \in \R$ is time and ${\bf s} \in \T^\mathnormal{d}$ the spatial coordinate vector.
Take a spatial test function $c({\bf s})$ and define the temporal process $\Omega_1(t)$ 
formally as 
\begin {equation}
\label {wn_t1}
\Omega_1(t) := \int_{\T^\mathnormal{d}} \Omega(t, {\bf s}) c({\bf s})  \,\d {\bf s},
\end {equation}
so that it acts on a test function in the temporal domain, $\varphi(t)$, as 
\begin {equation}
\label {wn_t1a}
(\Omega_1, \varphi):= \int_{\R} \Omega_1(t) \varphi(t)\,\d t =
  \int_{\R} \int_{\T^\mathnormal{d}} \Omega(t, {\bf s}) c({\bf s}) \varphi(t) \,\d {\bf s} \,\d t.
\end {equation}
Here, we note that the latter double integral is nothing other than the result of action
of the original white noise $\Omega(t, {\bf s})$ on the spatio-temporal test function
$c({\bf s}) \cdot \varphi(t)$.
This enables us to mathematically rigorously {\em define} $\Omega_1(t)$ 
as the generalized random process that, with the fixed $c({\bf s})$,
 acts on the test function $\varphi(t)$ as follows:
\begin {equation}
\label {wn_t1b}
(\Omega_1(t), \varphi(t)):= (\Omega(t,{\bf s}), \,c({\bf s})  \varphi(t)).
\end {equation}
Now, using the definition Eq.(\ref{wn2}) of the white noise $\Omega(t, {\bf s})$, we have
\begin {equation}
\label {wn_t3}
\Ex|(\Omega(t, {\bf s}), \,c({\bf s}) \varphi(t))|^2 =
    \int_{\T^\mathnormal{d}}  \int_{\R} |c({\bf s})|^2 |\varphi(t)|^2 \,\d {\bf s} \,\d t =
    \int_{\T^\mathnormal{d}} |c({\bf s})|^2 \,\d {\bf s}  \int_{\R}  |\varphi(t)|^2 \,\d t.
\end {equation}
Since we have fixed $c({\bf s})$, we observe that
\begin {equation}
\label {wn_t3a}
\sigma^2 := \int |c({\bf s})|^2 \,\d {\bf s}
\end {equation}
is a constant such that
\begin {equation}
\label {wn_t4}
\Ex|(\Omega_1(t), \varphi(t))|^2 = \sigma^2  \int_{\R}  |\varphi(t)|^2 \,\d t.
\end {equation}
Comparing this equation with one of the definitions of the standard white noise, Eq.(\ref{wn2}), 
we recognize $\Omega_1(t)$ as a general Gaussian white noise in time,
\ie the standard temporal white noise multiplied by $\sigma$.
We call $\sigma$ the {\em intensity} of the white noise.

\subsection{Spatial spectrum of a spatio-temporal white noise}
\label{sec_wnoise_spaspe}

Now, we are in a position to derive the spatial spectrum of the standard
spatio-temporal white noise $\Omega(t, {\bf s})$.
In the formal Fourier decomposition
\begin {equation}
\label {wns1}
\Omega(t, {\bf s})=\sum_{\bf k} \tilde\Omega_{\bf k}(t) \,\e^{\i ({\bf k}, {\bf s})},
\end {equation}
the elementary temporal processes $\tilde\Omega_{\bf k}(t)$ can be shown to be white noises in time.
Indeed, again formally, we have
\begin {equation}
\label {wns2}
\tilde\Omega_{\bf k}(t) = \frac{1}{(2\pi)^d} 
     \int_{\T^\mathnormal{d}} \Omega(t, {\bf s}) \,\e^{-\i ({\bf k}, {\bf s})} \,\d {\bf s}.
\end {equation}
Here, we recognize an expression of the kind given by Eq.(\ref{wn_t1})
with $c({\bf s}):= \e^{-\i ({\bf k}, {\bf s})} / (2\pi)^d$.
Therefore, from Eq.(\ref{wn_t4}), $\tilde\Omega_{\bf k}(t)$
is a temporal white noise with the intensity $\sigma_{\bf k}^\Omega$ squared equal to 
\begin {equation}
\label {wns3}
(\sigma_{\bf k}^\Omega)^2 = \int |c({\bf s})|^2 \,\d {\bf s} =  \frac{1}{(2\pi)^{2d}} 
     \int_{\T^\mathnormal{d}} |\e^{-\i ({\bf k}, {\bf s})}|^2 \,\d {\bf s} = \frac{1}{(2\pi)^d}.
\end {equation}
In addition, using Eq.(\ref{wn3}), it is easy to show that $\tilde\Omega_{\bf k}(t)$ and
$\tilde\Omega_{\bf k'}(t)$ are mutually orthogonal for ${\bf k} \ne {\bf k}'$.

To summarize, $\tilde\Omega_{\bf k}(t)$ are 
mutually orthogonal  white-in-time noises, all with equal intensities 
$\sigma_{\bf k}^\Omega=(2\pi)^{-d/2}$:
\begin {equation}
\label {wns3a}
 \tilde\Omega_{\bf k}(t) =  \frac{1}{(2\pi)^{d/2}} \,\Omega_{\bf k}(t),
\end {equation}
where $\Omega_{\bf k}(t)$ are the standard white noises.

Note that if we consider the non-unit torus 
$\T^\mathnormal{d}(R)=\S^1(R) \times \dots \times \S^1(R)$ ($d$ times), where
$R$ is the radius of each circle, then, obviously, in Eqs.(\ref{wns2})--(\ref{wns3a}),
$2\pi$ is to be replaced by $2\pi R$.

\subsection{Spectral decomposition of a white in time and colored in space noise}
\label{sec_sigma_k}

In order to introduce a white in time and colored in space noise, let us 
{\em convolve} the spatio-temporal white noise  $\Omega(t, {\bf s})$ with 
 a {\em smoothing kernel} in space $u({\bf s})$, getting
\begin {equation}
\label {al1}
\alpha(t, {\bf s}) := \int_{\T^\mathnormal{d}} u({\bf s-r}) \,\Omega(t, {\bf r})  \,\d {\bf r}.
\end {equation}
In this equation, the stochastic integral is defined, for any $t$ and ${\bf s}$, following 
Eq.(\ref{wn_t1b}) with $c({\bf r}) := u({\bf s-r})$.
Fourier transforming $u({\bf s})$,
\begin {equation}
\label {al2}
u({\bf s})=\sum_{\bf k} \tilde u_{\bf k} \,\e^{\i ({\bf k}, {\bf s})},
\end {equation}
and, in space, $\alpha(t, {\bf s})$,
\begin {equation}
\label {al3}
\alpha(t, {\bf s}) = \sum_{\bf k} \tilde\alpha_{\bf k}(t) \,\e^{\i ({\bf k}, {\bf s})},
\end {equation}
we easily obtain that the elementary spectral processes $\tilde\alpha_{\bf k}(t)$
are independent white noises in time with the intensities squared
\begin {equation}
\label {al4}
\sigma_{\bf k}^2=(2\pi)^{d} |\tilde u_{\bf k}|^2,
\end {equation}
so that the stochastic differential $\tilde\alpha_{\bf k}(t) \d t$ is
\begin {equation}
\label {al5}
\tilde\alpha_{\bf k}(t) \,\d t =  \sigma_{\bf k} \,\d W_{\bf k}(t).
\end {equation}
Equivalently, 
\begin {equation}
\label {al6}
\tilde\alpha_{\bf k}(t)  =  \sigma_{\bf k} \,\Omega_{\bf k}(t).
\end {equation}
%

\subsection{Discretization of the spectral processes $\tilde\alpha_{\bf k}(t)$ in time}
\label{sec_dscr_wnoise}

Being white noises, $\tilde\alpha_{\bf k}(t)$ have infinite variances.
They become ordinary random processes if, e.g., we discretize them in time.
With the time step $\Delta t$, we define the discretized process 
$\hat\alpha_{\bf k}(t_j)$ at the time instance $t_j$ 
by replacing, in Eq.(\ref{al5}),  $\d t$ with $\Delta t$ and $\d W_{\bf k}$ with $\Delta W_{\bf k}$:
\begin {equation}
\label {ad1}
\hat\alpha_{\bf k}(t_j) \,\Delta t :=  \sigma_{\bf k}  \,\Delta W_{\bf k}(t).
\end {equation}
As $\Ex |\Delta W_{\bf k}(t)|^2 = \Delta t$, we  obtain
\begin {equation}
\label {ad2}
\hat\alpha_{\bf k}(t_j) =   \frac{\sigma_{\bf k}} {\sqrt{\Delta t}}     \cdot \zeta_{{\bf k}j},
\end {equation}
where  $\zeta_{{\bf k}j}$ are independent complex standard Gaussian random variables $CN(0,1)$.
The latter is defined as a complex random variable whose
 real and imaginary parts are mutually uncorrelated zero-mean random variables with variances equal to $1/2$.
 $CN(0,1)$ is  sometimes  referred to as the circularly symmetric complex Gaussian (normal) random variable \citep[e.g.][]{Tse}.

Equation (\ref{ad2}) shows that the spatial spectrum of the time discrete driving noise is 
$\sigma_{\bf k}^2 / \Delta t$.

\section{Physical-space approximation of the  operator $\sqrt{1 - \lambda^2\Delta}$}
\label{app_symbol}

As we have discussed in section \ref{sec_PSproperty}, the  fractional power (square root)
of the negated and shifted  Laplacian operator, 
${\cal L} := \sqrt{1 - \lambda^2\Delta}$,
is defined as the pseudo-differential operator  with the symbol
$\tilde l({\bf k}) := \sqrt{1 + \lambda^2{\bf k}^2}$.
In the literature, one can find approaches to discretization
of fractional powers of elliptic operators, \eg \citet[][]{Simpson} used finite elements in the spatial context.

Here, we propose a  simple technique to build a spatial discretization scheme
that approximates the  operator $\sqrt{1 - \lambda^2\Delta}$ in the sense
that the symbol of the approximating operator is close
to $\sqrt{1 + \lambda^2{\bf k}^2}$.

To this end, we do the following.
\begin{enumerate}
\item
 Perform the backward Fourier transform of the symbol $\tilde l({\bf k})$, getting the function $l({\bf s})$. 
 As multiplication in Fourier space by  $\tilde l({\bf k})$ is equivalent to convolution in physical space with $l({\bf s})$, 
 we obtain that for any test function $\varphi({\bf s})$,
\begin {equation}
\label {fL1}
({\cal L}\varphi)({\bf s}) = \int_{\T^3}  l({\bf s-r}) \, \varphi({\bf r}) \,\d {\bf r}.
\end {equation}
The crucial moment here is that the kernel function $l({\bf s})$ appears to be oscillating while rapidly decreasing
in modulus as $|{\bf s}|$ increases (see below). 
This enables its efficient approximation with a compact-support  (truncated) function.

\item
With the discretization on the grid with $n$ points in each of the $d$ dimensions
on the torus $\T^\mathnormal{d}$, the kernel function $l({\bf s})$ is represented by the set of its grid-point values
 $l({\bf s}_{\bf i})$, 
 where 
  ${\bf s}=(s_1,\dots, s_d)$,
 ${\bf i}=(i_1,\dots, i_d)$, and 
 ${\bf s}_{\bf i}=(s_1({i_1}),\dots, s_d({i_d}))$.
If $l({\bf s})$ appears to be rapidly decreasing away from ${\bf s=0}$, we 
truncate the $l({\bf s}_{\bf i})$ function by limiting its support near the origin, thus
getting the function $l_{\rm trunc}({\bf s}_{\bf i})$.
\eeg in  3D, the support of $l_{\rm trunc}({\bf s}_{\bf i})$
consists of the grid points ${\bf i}=(i_1, i_2, i_3)$
that simultaneously satisfy the following constraints:
$|i_1| \le J$, 
$|i_2| \le J$, and
$|i_3| \le J$, 
 where $J$ is the spatial order of the scheme.
 Below, we present results with $J=1$ (3 grid points in the support of the 
 truncated kernel function in each dimension)
 and $J=3$ (7 grid points in the support in each dimension).

\item
Fourier transform $l_{\rm trunc}({\bf s})$ back to the spectral space,
getting the approximated symbol $\tilde l_{\rm trunc}({\bf k})$.

\item
Compare $\tilde l({\bf k})$ with $\tilde l_{\rm trunc}({\bf k})$ and conclude 
whether a parsimonious (that is, with a very small $J$)  approximation is viable.
  
\end{enumerate}
  
Now, we present the results.
We found that for $d=1$, $d=2$, and $d=3$, the goodness of fit 
was similar, so we examine the 3D case below.

We selected the grid of $n = 2\cdot n_{\rm max}=256$ points in each of the three dimensions. 
We specified the  spatial non-dimensional length scale $\lambda$ to be much greater than the mesh size $h=2\pi/n$ and much less than the domain's extents,
$2\pi$. Specifically, we chose $\lambda=1/n_1$, where $n_1:=\sqrt{n_{\rm max}}$. 
(The results were not much sensitive 
to changes in $n_1$ within the whole wavenumber range on the grid.)
 
\begin{figure}
\begin{center}
   { \scalebox{0.4}{ \includegraphics{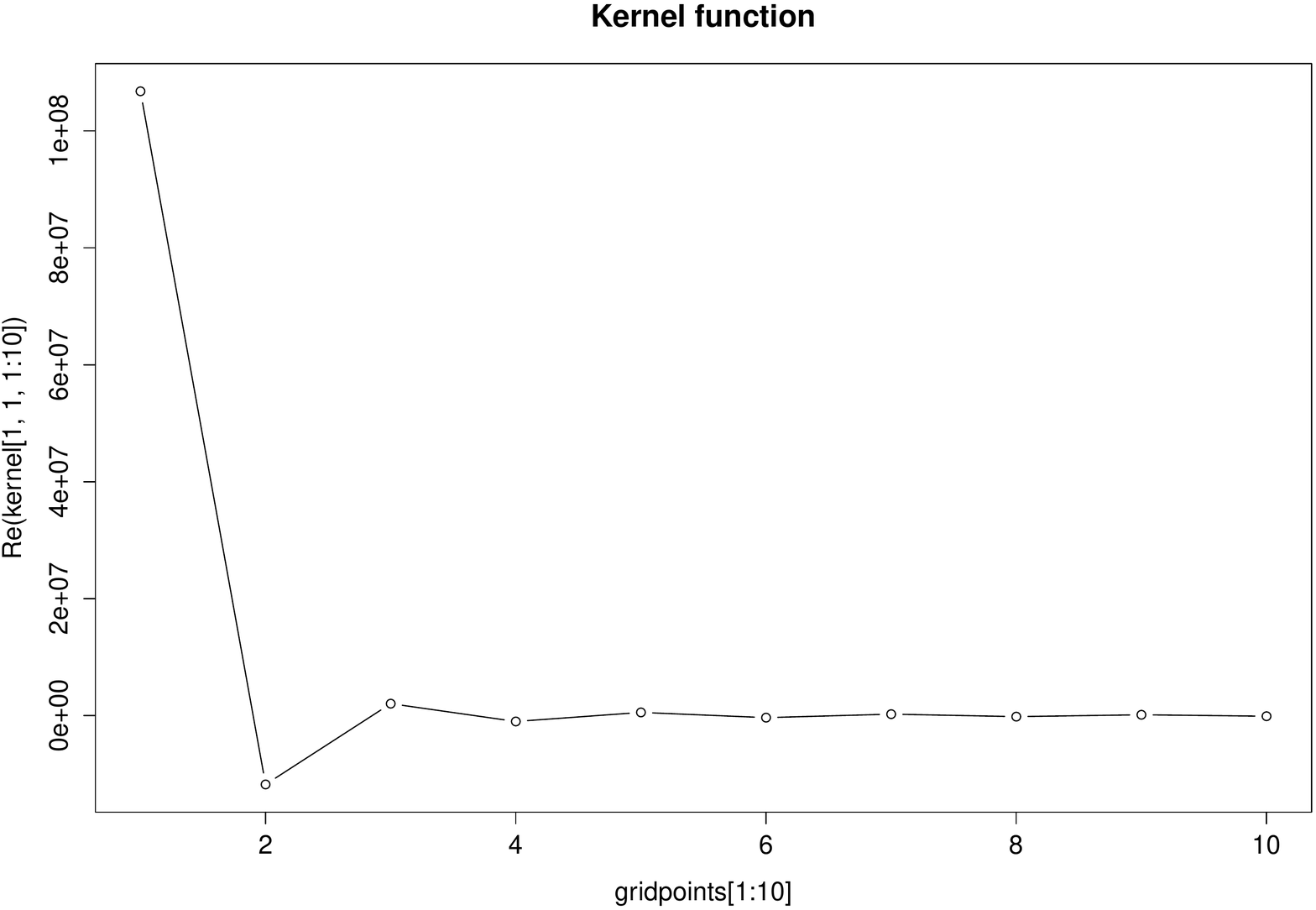}}
       }
\end{center}
  \caption{The kernel function}
\label{Fig_kernel}
\end{figure}

 Figure \ref{Fig_kernel} displays the resulting kernel function $l({s})$ for positive $s$
 (note that $l({s})$ is an even function of the scalar distance $s$).
 One can see the remarkably fast decay of $|l({s})|$ with the growing $s$.
Consequently, a stencil with just a few points in each dimension 
can be expected to work well.

\begin{figure}
\begin{center}
   { \scalebox{0.26}{ \includegraphics{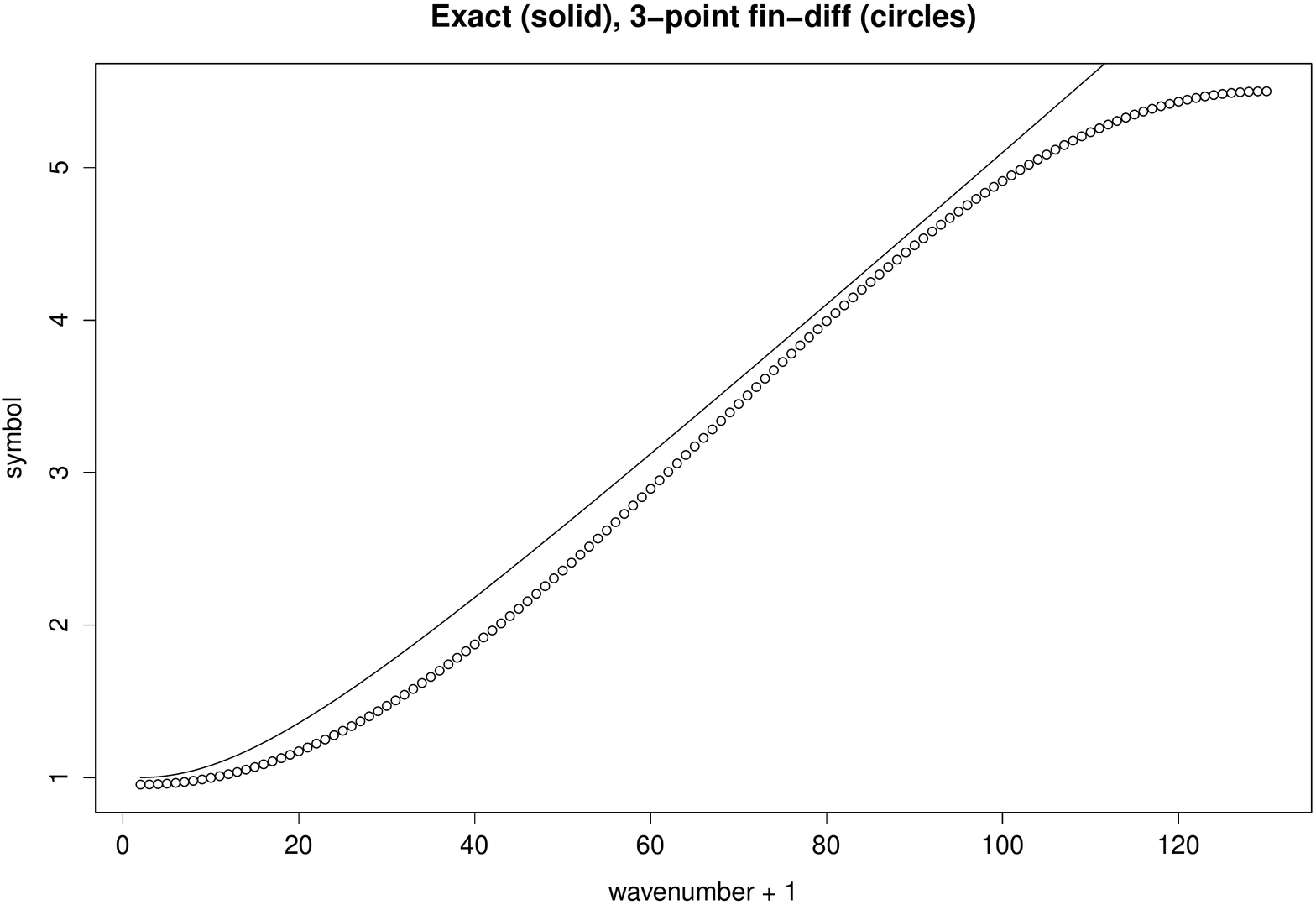}}
     \scalebox{0.26}{ \includegraphics{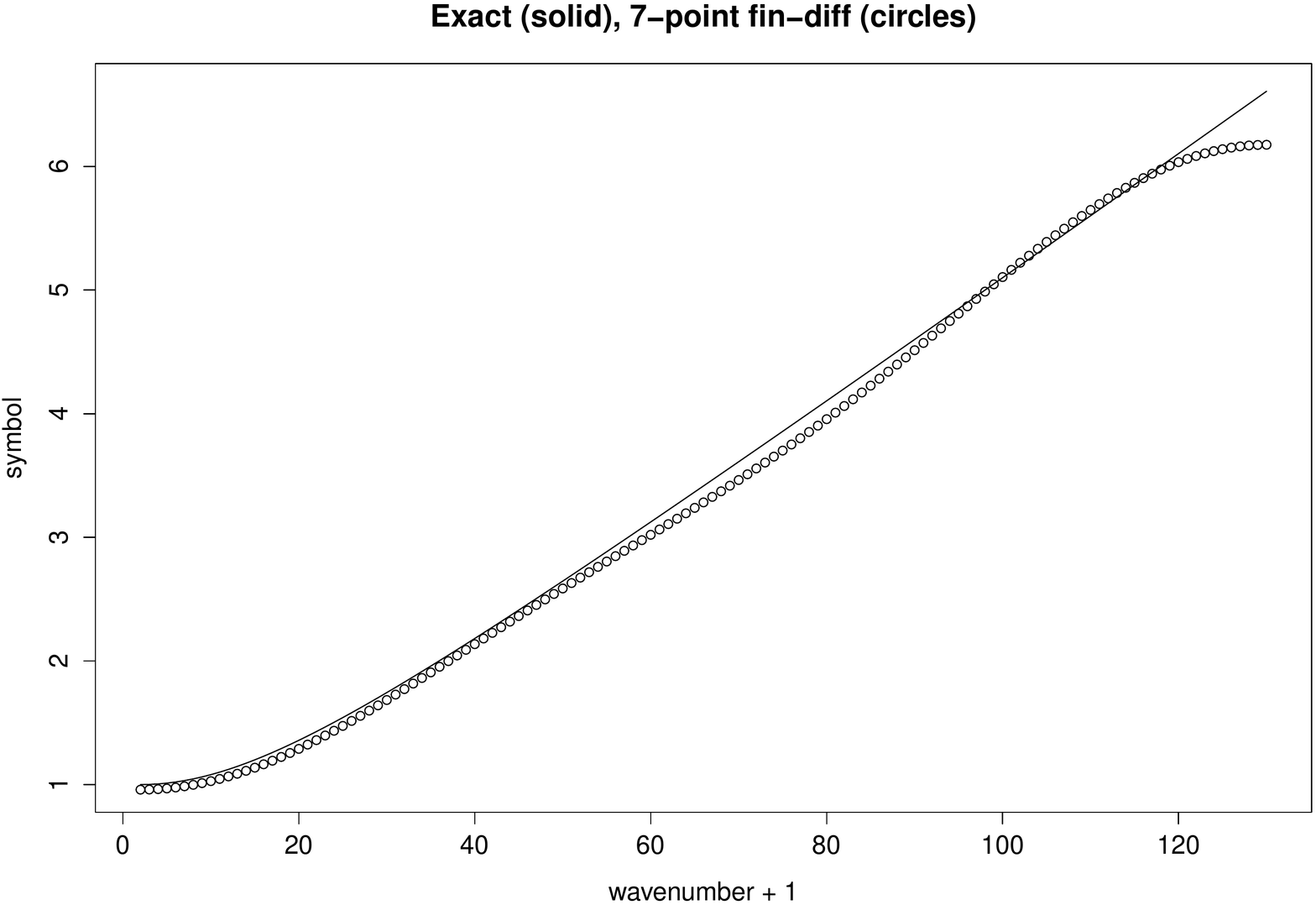}}
       }
\end{center}
  \caption{Goodness of fit of the symbol $\tilde l({\bf k})$ (solid curve) 
  by $\tilde l_{\rm trunc}({\bf k})$ (circles). 
  Left: the 3-point stencil in each dimension.
    Right: the 7-point stencil}
\label{Fig_symbol}
\end{figure}

 Figure \ref{Fig_symbol} shows the exact and approximated symbols for the
 stencil that contains 3 grid points in each dimension (the left panel) and
the  stencil that contains 7 grid points in each dimension (the right panel).
  (The  5-point scheme worked not much better than the 3-point one and so its performance
  is not shown.)

From Fig.\ref{Fig_symbol}, one can see that the 3-point scheme's performance is rather
mediocre, whereas the 7-point scheme works very well (in terms of the reproduction of the 
operator's symbol).

Finally, we verified that the symbol $\tilde l_{\rm trunc}({\bf k})$ of the discrete operator for $J=3,5,7$ was everywhere positive,
which guarantees that the operator is positive definite and so the discretized SPG model
should be stable.

To summarize, the operator $\sqrt{1 - \lambda^2\Delta}$
can be approximated with parsimonious physical-space discretization schemes.
For simulation of uncertainty in meteorology, where precise error statistics is not 
available, the simplest 3-point (in each direction) scheme seems most appropriate
and computationally attractive.
For more demanding applications, the 7-point scheme can be more appropriate.

\section{Stationary statistics of a higher-order OSDE}
\label{app_OSDE}

We examine the OSDE  Eq.(\ref{bpg_ps}) in its generic form:
\begin{linenomath*}
\begin {equation}
\label {osde_p}
\left(\frac{\d} {\d t} +  a\right)^p \eta(t) =  \sigma \Omega(t),
\end {equation}
\end{linenomath*}
where $\eta(t)$ is the random process in question, 
$\sigma$ and $a$ are the positive numbers, $p$ is the positive integer, and 
$\Omega(t)$ is the standard white noise,
see \eg \citet[][section 1.1.3]{Rozanov} or \citet[section 3.1.4]{Kuo}
and also  Appendix \ref{app_wnoise_1}.

The goal here is to find the variance and the correlation function of $\eta(t)$ in the
stationary regime.
The technique is to reduce the $p$-th order OSDE to a system of first-order OSDEs.

To simplify the exposition, we consider the third-order OSDE ($p=3$) and rewrite Eq.(\ref{osde_p}) as
\begin{linenomath*}
\begin {equation}
\label {osde_p1}
 \left(\frac{\d} {\d t} +  a\right) \,\,\,\,
 \left\{ \left(\frac{\d} {\d t} +  a\right) \,\,
 \left[ \left(\frac{\d} {\d t} +  a\right)   \eta(t) \right] \right\}=  \sigma \Omega(t).
\end {equation}
\end{linenomath*}
Here, by $\eta_1$ we denote the term in brackets,
\begin{linenomath*}
\begin {equation}
\label {osde_p2}
 \left(\frac{\d} {\d t} +  a\right) \eta =  \eta_1 
\end {equation}
\end{linenomath*}
and by $\eta_2$ the term in braces,
\begin{linenomath*}
\begin {equation}
\label {osde_p3}
 \left(\frac{\d} {\d t} +  a\right) \eta_1 =  \eta_2,
\end {equation}
\end{linenomath*}
so that  Eq.(\ref{osde_p1}) implies that
\begin{linenomath*}
\begin {equation}
\label {osde_p4}
 \left(\frac{\d} {\d t} +  a\right) \eta_2 =  \sigma \Omega.
\end {equation}
\end{linenomath*}
In Eqs.(\ref{osde_p2})--(\ref{osde_p4}), the last equation is the familiar first-order OSDE forced by the white noise, 
whereas the other equations are not forced by the white noise.
Generalizing the above construction, Eqs.(\ref{osde_p1})--(\ref{osde_p4}), to the arbitrary $p>0$,
we  form the following first-order vector-matrix OSDE (a system of first-order OSDEs):
\begin{linenomath*}
\begin {equation}
\label {osde_p5}
\d\boldsymbol\eta  +  {\bf A} \boldsymbol\eta \d t=  \boldsymbol\Sigma \boldsymbol\Omega \d t,
\end {equation}
\end{linenomath*}
where $\boldsymbol\eta = ( \eta,  \eta_1, \dots,  \eta_{p-2},  \eta_{p-1})$, 
$\boldsymbol\Omega= ( 0,  0, \dots,  0, \Omega)$ , and the design of the matrices 
${\bf A}$ and $\boldsymbol\Sigma$ is obvious (not shown).

With Eq.(\ref{osde_p5}) in hand, we derive a differential equation for the covariance matrix 
${\bf P} = \Ex \boldsymbol\eta \boldsymbol\eta^*$,
where $^*$ denotes transpose complex conjugate  \citep[e.g.][example 4.16]{Jazwinski}.
First, we compute the increment of ${\bf P}$:
\begin{linenomath*}
\begin {equation}
\label {dP}
\Delta {\bf P} =  \Ex  (\boldsymbol\eta + \d\boldsymbol\eta) (\boldsymbol\eta + \d\boldsymbol\eta)^* -  
              \Ex  \boldsymbol\eta \boldsymbol\eta^* = 
      \Ex \boldsymbol\eta\d\boldsymbol\eta^* +   \Ex \d\boldsymbol\eta\boldsymbol\eta^* +  
      \Ex \d\boldsymbol\eta\d\boldsymbol\eta^*.
\end {equation}
\end{linenomath*}
Then, using Eq.(\ref{osde_p5}) and the fact that $\Ex |\Omega \d t|^2 = \Ex |\d W|^2=\d t$, 
we obtain the differential of ${\bf P}$  from  Eq.(\ref{dP}): 
\begin{linenomath*}
\begin {equation}
\label {dP2}
\d {\bf P} =  -{\bf A}{\bf P} \d t - {\bf P}{\bf A}^*\d t +   \boldsymbol\Sigma \boldsymbol\Sigma^* \d t.
\end {equation}
\end{linenomath*}
In the stationary regime $\d {\bf P} =0$, so  the  equation for the stationary covariance matrix is
\begin{linenomath*}
\begin {equation}
\label {dP3}
{\bf A}{\bf P} + {\bf P}{\bf A}^* = \boldsymbol\Sigma \boldsymbol\Sigma^*.
\end {equation}
\end{linenomath*}
This a system of linear algebraic equations for the unknown entries of the matrix ${\bf P}$.
Because both ${\bf P}$ and $\boldsymbol\Sigma \boldsymbol\Sigma^*$ are self-adjoint matrices,
the number of unknowns, $p(p+1)/2$, is equal to the number of independent equations.
We analytically solve this system of equations and 
look at the {\em first} diagonal entry of the solution ${\bf P}$,
which represents the required $\Var\eta$ (because the random field in question
 $\eta$ is defined above to be 
the first entry of the vector $\boldsymbol\eta$).
Dropping tedious derivations, we present in Table \ref{Tab_osde} (the second row) 
the formulas for the temporal orders  $p=1$, $p=2$,  $p=3$, and
for the general $p$.

Finally, we derive the temporal correlation function for the $p$th-order OSDE.
To this end, we multiply Eq.(\ref{osde_p}) by $\eta(s)$ with $s <t$ and take expectation.
Since $a$ is non-stochastic, we may interchange the expectation and the differential operator 
$\left(\frac{\d} {\d t} +  a\right)^p$, getting the $p$th-order ordinary differential equation for 
the temporal covariance function, whose solutions for different $p$ are presented in row 3
of  Table \ref{Tab_osde}.

\begin{table}
\caption{\bf Variances $\Var\eta$ and correlation functions $C_\eta(t)$ of the stationary solution to Eq.(\ref{osde_p})
         for different temporal orders $p$.}

\begin{center}
\begin{tabular}{|l||c|c|c|c|}

\hline
  $p$    & $1$  & $2$ &  $3$ & Arbitrary $p$\\
      \hline
  $\Var\eta$ &   $\frac{\sigma^2}{2a}$ &  $\frac{\sigma^2}{4a^3}$   & 
     $\frac{3\sigma^2}{16a^5}$ & $\frac{\sigma^2}{a^{2p-1}}$ \\
  \hline
   $C_\eta(t)$    &   $\e^\mathnormal{-a|t|}$          &  $(1 + a|t|) \,\e^\mathnormal{-a|t|}$ & 
     $(1 + a|t| + \frac{a^2t^2} {3}) \,\e^\mathnormal{-a|t|}$ & $R_{p-1}(a|t|) \cdot \e^\mathnormal{-a|t|}$ \\
\hline
\end{tabular}
\end{center}
\small{\em Note that $R_{p-1}(.)$ in the last row of the table stands for the polynomial of order $p-1$.}
\label{Tab_osde}
\end{table}

\section{Smoothness of sample paths of the spatial Mat\'ern random field for different $\nu$}
\label{app_smoo}

\begin{figure}
\begin{center}
   { \resizebox{140mm}{55mm}{ \includegraphics{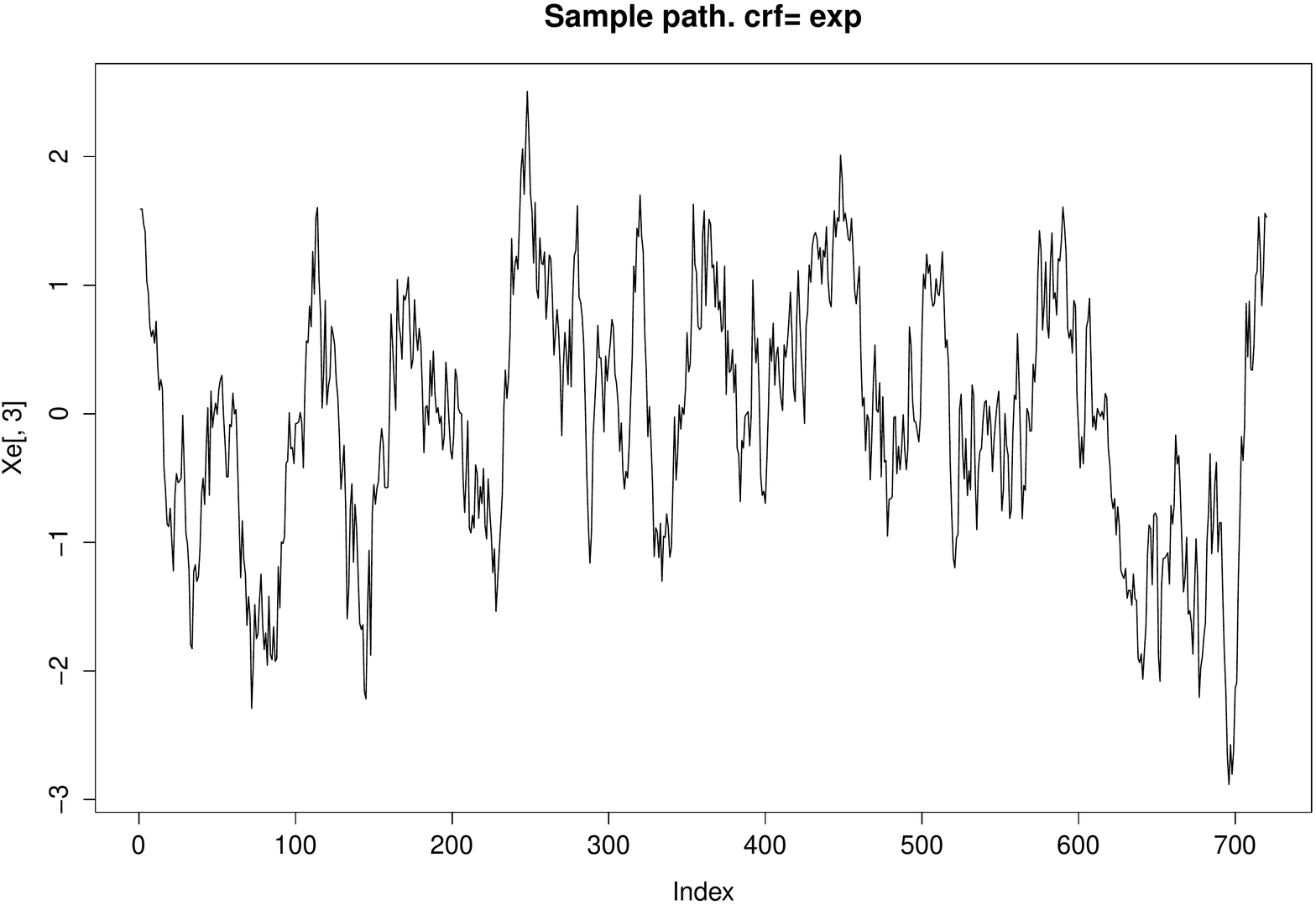}}
   
    \resizebox{140mm}{55mm}{ \includegraphics{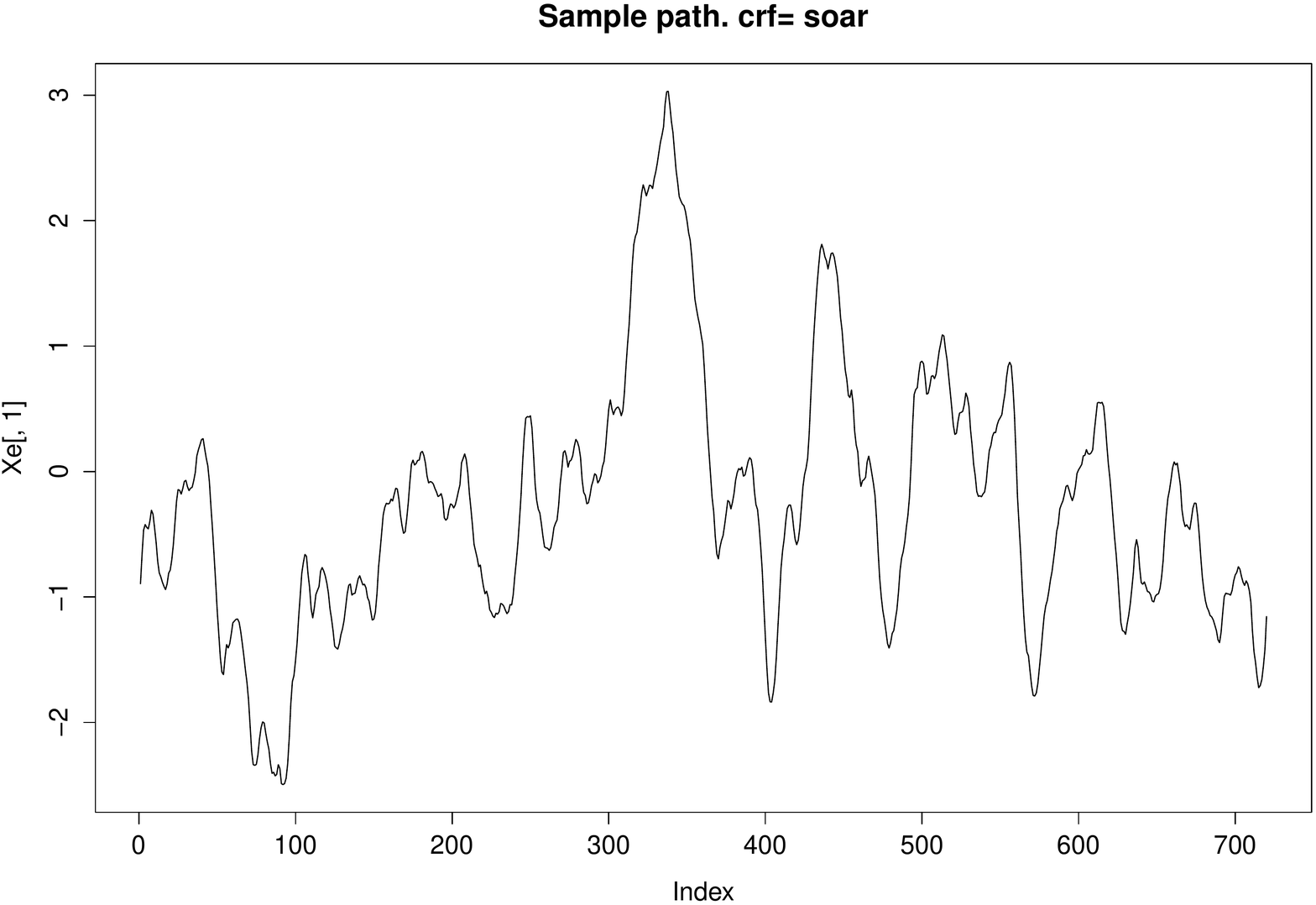}}
   
    \resizebox{140mm}{55mm}{ \includegraphics{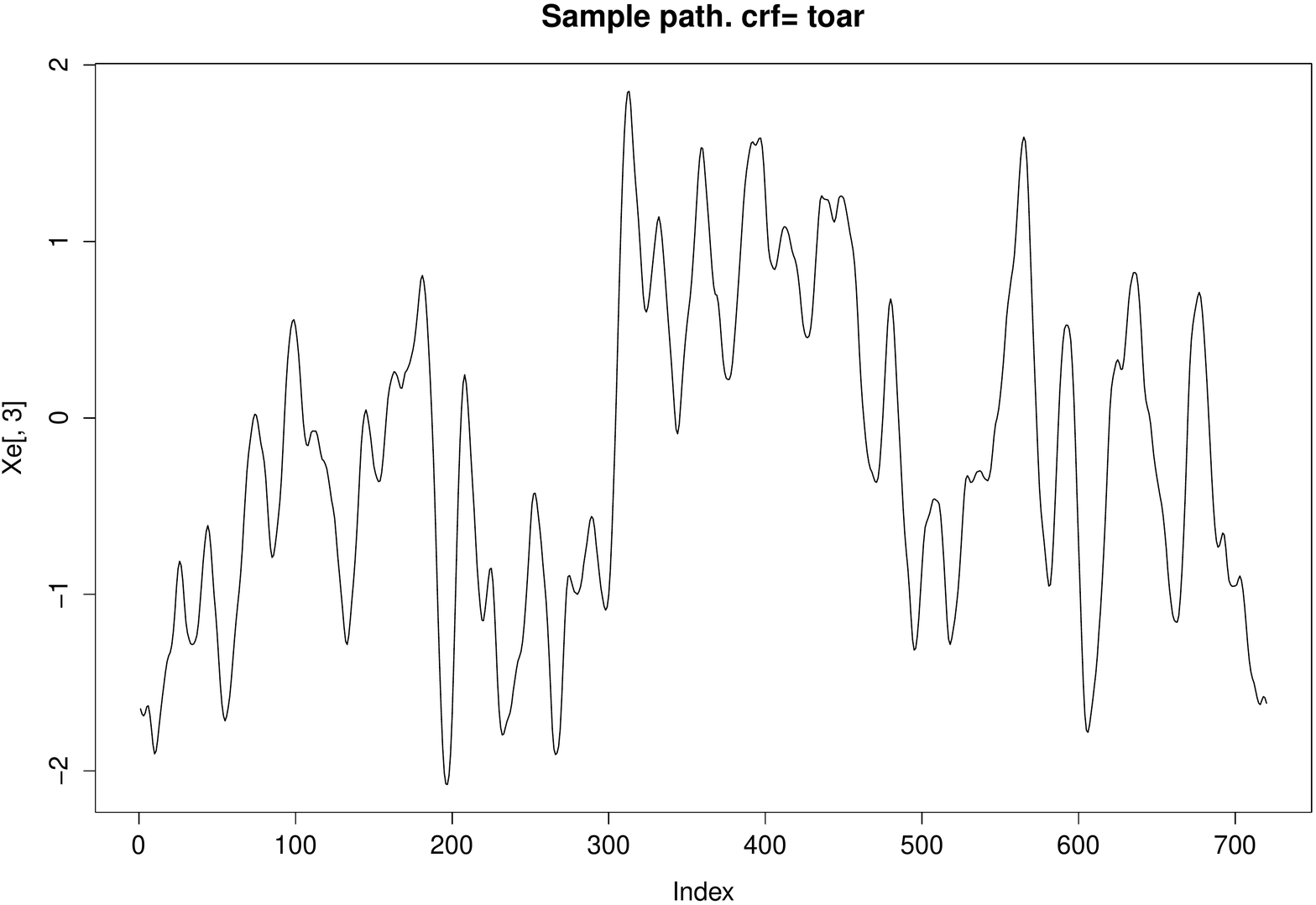}}
   
    \resizebox{140mm}{55mm}{ \includegraphics{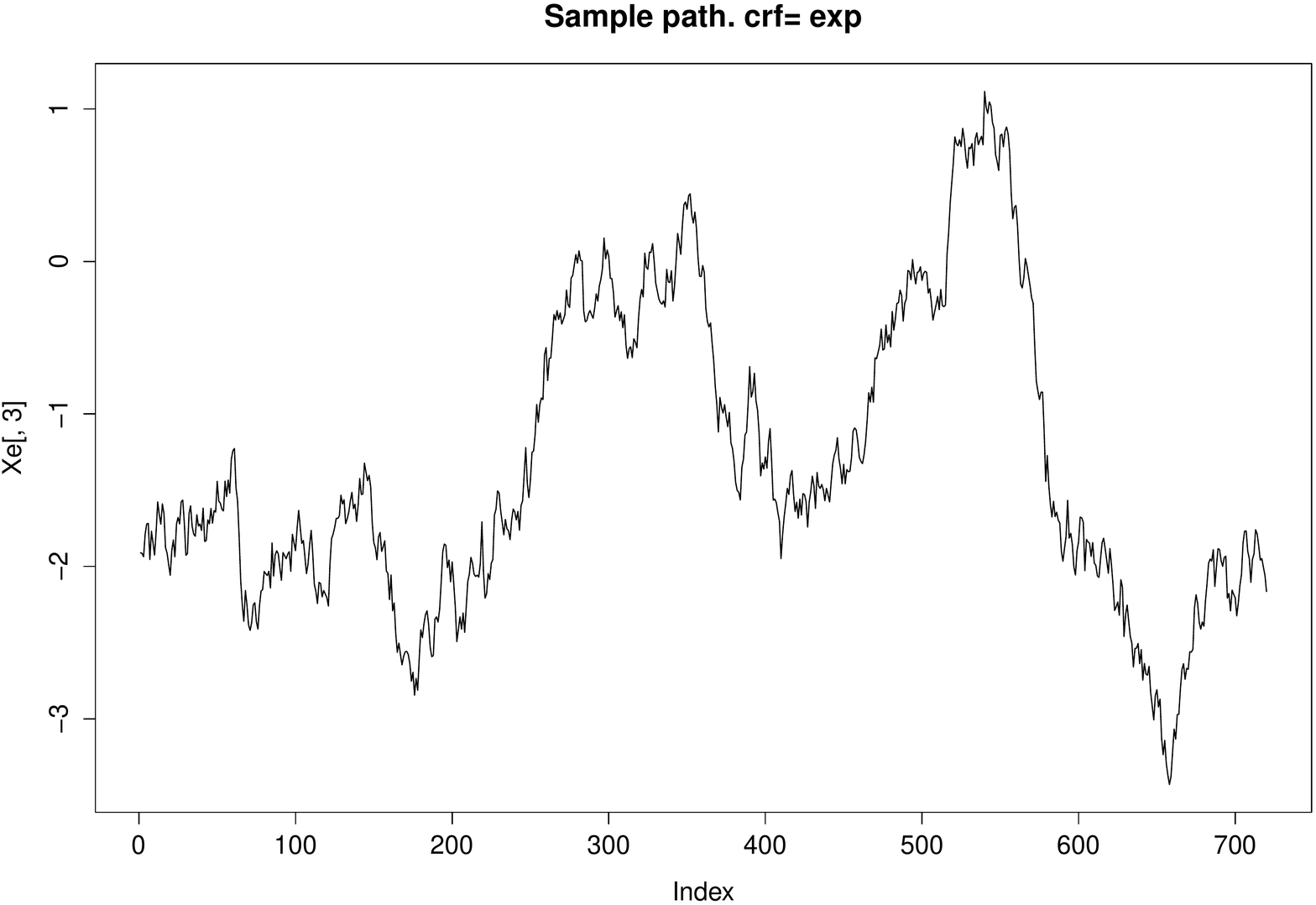}}
       }
\end{center}
  \caption{\small Sample paths for various $\nu$ and $L_{0.7}$. From the {\em top} to the {\em bottom}: \\
  ($\nu=\frac12$, $L_{0.7}=500$), 
  ($\nu=\frac32$, $L_{0.7}=500$), 
  ($\nu=\frac52$, $L_{0.7}=500$), 
  ($\nu=\frac12$, $L_{0.7}=1500$)
  }
\label{Fig_1D_realiz}
\end{figure}

Here, we show how sample paths (realizations) of the  Mat\'ern random field
with the smoothness parameter $\nu$ look.
Specifically, in Fig.\ref{Fig_1D_realiz},
we present three plots with 1D cross-sections of randomly chosen realizations 
of the Mat\'ern random field
for the following three values of $\nu$:
1/2, 3/2, and 5/2. The spatial length scale parameter $\lambda$ is selected in each 
of the three cases in such a way that the spatial correlation function intersects the 
0.7 level at approximately the same distance (we denote this distance
by $L_{0.7}$): $L_{0.7}=500$ km.
Note that, as in section \ref{sec_spatim_illu},
we assume, for convenience, that the extent of the spatial domain in each
coordinate direction is 3000 km (rather than $2\pi$).
For comparison, we also display a  realization with  $L_{0.7}=1500$ km for $\nu=1/2$
(the bottom panel of Fig.\ref{Fig_1D_realiz}). 

One can see that, indeed, the larger $\nu$, the smoother the realizations---in the sense that
they have less small-scale ``noise''.
By contrast,  increasing the  length scale $\lambda$ (compare the top and bottom panels of Fig.\ref{Fig_1D_realiz})
makes the {\em large-scale} pattern smoother but does {\em not} remove the smallest scales.
So, the {\em large-scale} behavior is determined by the length scale $\lambda$, whereas the degree of 
{\em small-scale} smoothness/roughness depends predominantly on the smoothness parameter $\nu$.

\section{Stationary statistics of a time discrete higher-order OSDE}
\label{app_OSDE_dscr}

Here, to simplify the exposition, we first examine the simplest first-order (\ie with $p=1$) OSDE and then give the results for 
the third-order OSDE used in the current version of the SPG.

\subsection{First-order numerical scheme}

Discretization of the Langevin Eq.(\ref{osde2}) by an implicit scheme yields
\begin {equation}
\label {sch_p1}
\eta_i - \eta_{i-1} + a \eta_i \,\Delta t= \sigma \Delta W_i,
\end {equation}
so that
\begin {equation}
\label {sch_p1a}
\eta_i = \frac{\eta_{i-1} + \sigma \Delta W_i } {1 + a \Delta t}.
\end {equation}
In the stationary regime, $\Var \eta_i =\Var \eta_{i-1}$, whence, bearing in mind that 
$\Var \Delta W_i = \Delta t$ and $\Delta W_i$ is independent on the values of $\eta$ for all time 
moments up to and including the moment $i-1$, we apply the variance operator to both sides of Eq.(\ref{sch_p1a}) and
obtain the stationary variance
\begin {equation}
\label {sch_p1V}
V(\Delta t) := \lim_{i \to \infty} \Var\eta_i = \frac{\sigma^2} { 2a + (g\Delta t)^2}.
\end {equation}
Note that, as $\Delta t \to 0$,  $V(\Delta t)$ tends to 
the continuous-time variance $\frac{\sigma^2} { 2a }$, 
see Eq.(\ref{osde4}).

\subsection{Third-order numerical scheme}

Consider the continuous-time OSDE, Eq.(\ref{osde_p}), with $p=3$.
The implicit scheme  Eq.(\ref{osde_num2}) we use to numerically solve it is reproduced here as
\begin {equation}
\label {sch_p3}
\eta_i = \frac{1} {\varkappa^3} 
   \left[ 3 \varkappa^2 \eta_{i-1} -3\varkappa\eta_{i-2}+\eta_{i-3}+ 
     \sigma  (\Delta t)^2 \,\Delta W_i \right],
\end {equation}
where  $\varkappa:=1+ a\Delta t$.
Here, the goal is to find the stationary variance $V := \lim_{i \to \infty} \Var\eta_i$ along with lag-1 and lag-2 stationary covariances,
$c_1 :=\lim_{i \to \infty} \Ex \eta_i \eta_{i-1}$ and 
$c_2 := \lim_{i \to \infty}\Ex \eta_i \eta_{i-2}$,
respectively.
To reach this goal, we build three linear algebraic equations for the 
three unknowns, $V$, $c_1$, and $c_2$.
The first equation is obtained by applying the variance operator to both sides of  Eq.(\ref{sch_p3}).
The second and third equations are obtained by multiplying Eq.(\ref{sch_p3}) by $\eta_{i-1}$ and $\eta_{i-2}$,  
respectively, and applying the expectation operator to both sides of the resulting  equations.
Omitting the derivations, we write down the results:
\begin {equation}
\label {sch_p3V}
V = \frac{\varkappa^4 + 4\varkappa^2 + 1} {(\varkappa^2 -1)^5} \,(\Delta t)^5 \sigma^2.
\end {equation}
\begin {equation}
\label {sch_p3c12}
c_1 = \frac{3\varkappa(\varkappa^2 + 1)} {(\varkappa^2 -1)^5} \,(\Delta t)^5 \sigma^2,
\qquad
       c_2 = \frac{6\varkappa^2} {(\varkappa^2 -1)^5} \,(\Delta t)^5 \sigma^2.
\end {equation}
As in the first-order case, one  can see that as $\Delta t \to 0$,  $V$ tends to 
the continuous-time variance $\frac{3}{16} \frac{\sigma^2} { a^5 }$, 
see Table \ref{Tab_osde}.


\bibliography{mybibfile}


\end{document}